\documentclass[aps,prb,superscriptaddress,twocolumn]{revtex4}
\usepackage{amssymb,amsmath,mathptmx}
\usepackage{graphicx}
\usepackage{dcolumn}
\usepackage{bm}
\usepackage{hyperref}
\usepackage{color}
\usepackage[utf8x]{inputenc}
\usepackage{array}
\newcolumntype{C}{>{\centering\arraybackslash}p{1em}}
\newcommand{\kbf}{\mathbf{k}}
\newcommand{\kpbf}{\mathbf{k'}}
\newcommand{\be}{\begin{equation}}
\newcommand{\ee}{\end{equation}}
\newcommand{\bea}{\begin{eqnarray}}
\newcommand{\eea}{\end{eqnarray}}
\begin{document}


\title{Berry phase in superconducting multiterminal
  quantum dots}

\author{Benoît Douçot}

\affiliation{Laboratoire de Physique Théorique et Hautes Energies,
  Sorbonne Universit\'e and CNRS UMR 7589, 4 place Jussieu, 75252
  Paris Cedex 05, France}

\author{Romain Danneau} 

\affiliation{Institute of Nanotechnology,
  Karlsruhe Institute of Technology, D-76021 Karlsruhe, Germany}

\author{Kang Yang}

\affiliation{Laboratoire de Physique Théorique et Hautes Energies,
  Sorbonne Universit\'e and CNRS UMR 7589, 4 place Jussieu, 75252
  Paris Cedex 05, France}

\affiliation{Laboratoire de Physique des Solides, CNRS UMR 8502,
  Univ. Paris-Sud, Universit\'e Paris-Saclay F-91405 Orsay Cedex, France}

\author{Jean-Guy Caputo}

\affiliation{Laboratoire de Math\'ematiques, INSA de Rouen, Avenue de
  l'Universit\'e, F-76801 Saint-Etienne du Rouvray, France}

\author{R\'egis M\'elin}

\affiliation{Univ. Grenoble-Alpes, CNRS, Grenoble INP\thanks{Institute
    of Engineering Univ. Grenoble Alpes}, Institut NEEL, 38000
  Grenoble, France}

\date{\today}
\begin{abstract} 

We report on the study of the non-trivial Berry phase in superconducting multiterminal quantum dots biased at
commensurate voltages. Starting with the time-periodic Bogoliubov-de Gennes equations, we obtain a tight binding model
in the Floquet space, and we  solve these equations in the semiclassical limit. We observe that the
    parameter space defined by the contact transparencies and quartet
    phase splits into two components with a non-trivial Berry phase. We use the Bohr-Sommerfeld quantization to calculate the Berry phase.
    We find that if the quantum dot level sits at zero energy, then the Berry phase takes the values $\varphi_B=0$ or $\varphi_B=\pi$.
    We demonstrate that this non-trivial Berry phase can be observed by tunneling spectroscopy in the Floquet spectra.
    Consequently, the Floquet-Wannier-Stark ladder spectra of  superconducting multiterminal
    quantum dots are shifted by half-a-period if $\varphi_B=\pi$.
    Our numerical calculations based on Keldysh Green's functions show that this Berry phase spectral shift
    can be observed from the quantum dot tunneling density of states.
\end{abstract}

\maketitle
\section{Introduction}

The geometric phase is a general concept common to both classical and
quantum physics \cite{Berry1}. In a quantum system, the wave function
can accumulate a geometric phase, also called Berry
phase, following cyclic adiabatic evolution
around the phase space origin
\cite{Pancha,Longuet,Berry2,Bohm,Xiao}. Over the years, the Berry
phase has been extensively studied both theoretically and
experimentally \cite{Bohm,Xiao} as it can provide
  deep insight on fundamental problems in qubits
\cite{Falci,Makhlin,Jones,Leek}, topological insulators \cite{Hasan},
skyrmions \cite{Fert}, single and bilayer graphene
\cite{Shytov,Young,Varlet,Du}, molecular physics \cite{Resta},
Bose-Einstein {condensates \cite{Yao,Gao} to} cite
but a few.

\begin{figure}[htb]
  \begin{minipage}{.45\columnwidth}
    \includegraphics[width=\textwidth]{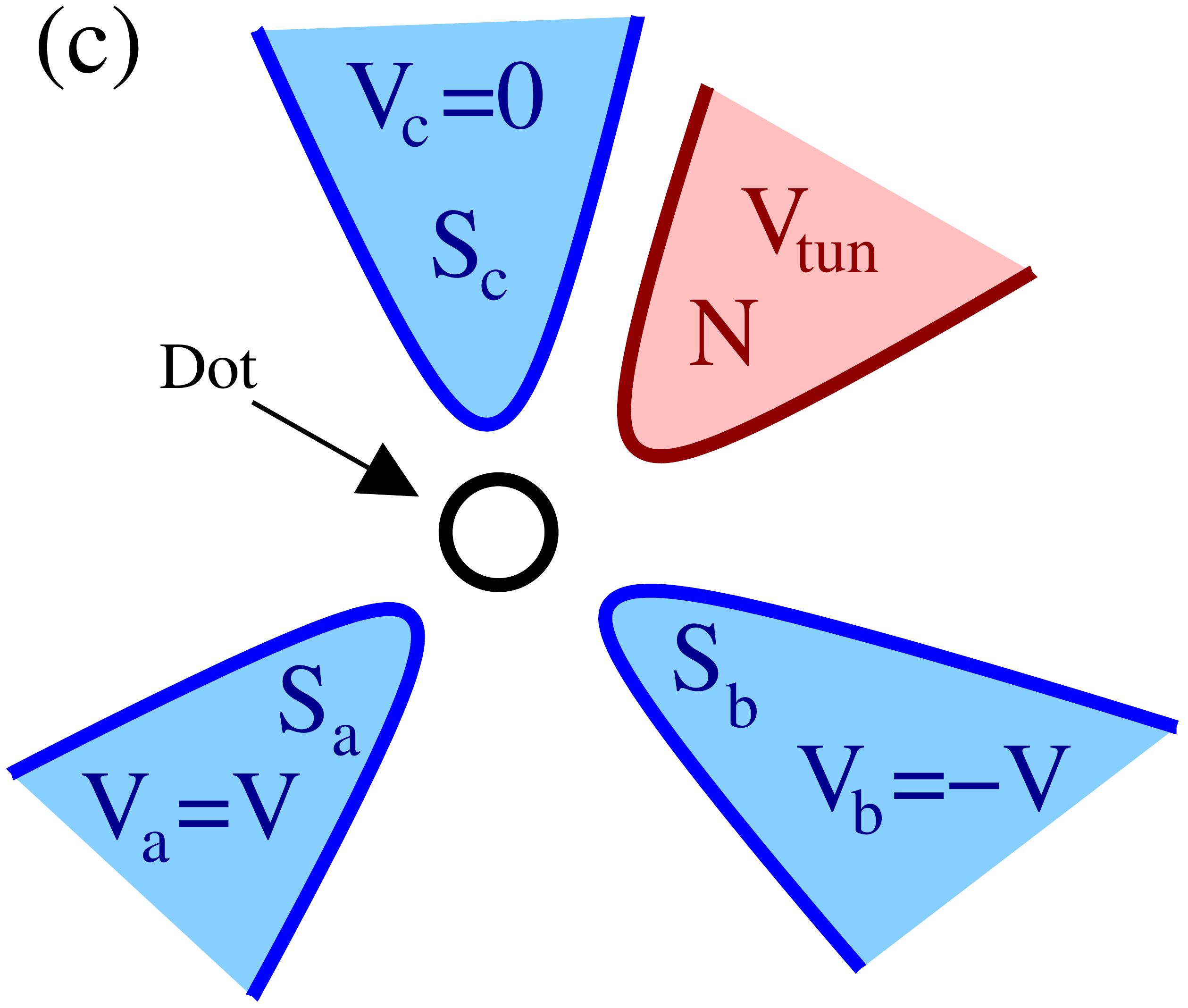}
  \end{minipage}
  \begin{minipage}{.54\columnwidth}
    \includegraphics[width=.9\textwidth]{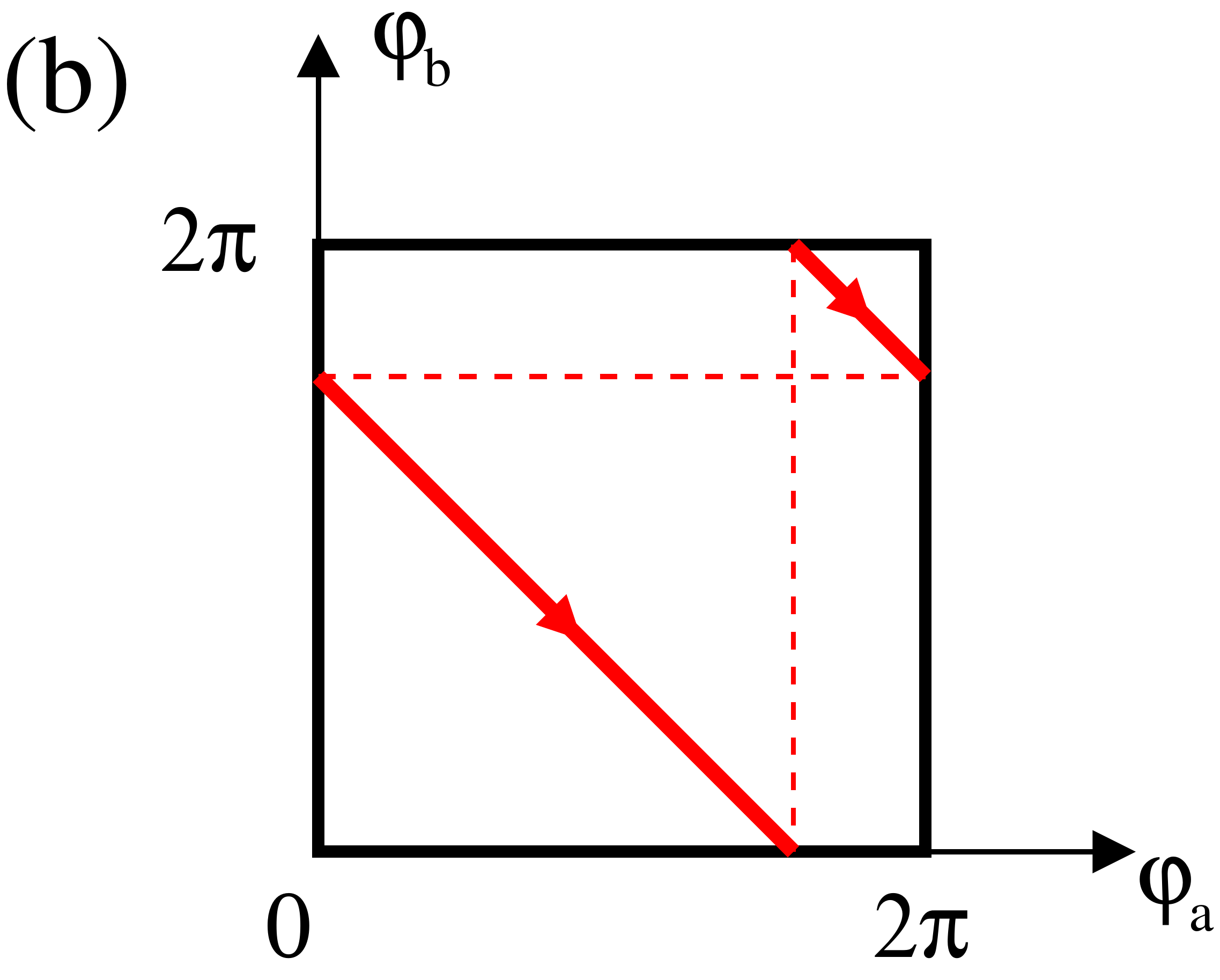}
  \end{minipage}

  \caption{(a) A superconducting three-terminal QD biased on the
    quartet line at voltages $V_{a,b}=\pm V$ and $V_c=0$,
    with in addition a tunnel-contacted normal
     lead to probe the quantum dot density of states. (b) The
    values of $(\varphi_a(t),\varphi_b(t))$ during one period of
    Josephson oscillations. Commensurate bias voltage implies that
    $(\varphi_a,\varphi_b)$ encloses a cycle on a two-dimensional
    torus. 
    \label{fig:0}
  }
\end{figure}

Recently, superconducting multiterminal devices have
triggered broad interest owing to many exotic
phenomena uncovered in these systems, like emergence of Majorana
fermions \cite{Badiane1,Houzet1,Badiane2}, topological states
associated to zero-energy Andreev Bound States
(ABS) and Weyl singularities
\cite{vanHeck,Padurariu,Riwar,Strambini,Eriksson},
  or new correlations among pairs of Cooper pairs so-called quartets
\cite{Freyn1,Jonckheere,Pfeffer,Melin1,Cohen}. As a new kind of
elementary process, the quartets appear when the leads
are driven by commensurate voltages in a three-terminal geometry (see
Fig.~\ref{fig:0}) and occur as the differential
  resistance features \cite{Pfeffer,Cohen} theoretically predicted in
Ref.~\onlinecite{Freyn1}. Moreover, in the case of superconducting
quantum dots (QD), we have lately demonstrated that
the nontrivial ABS time-periodic dynamics yields
sharp resonances in the Floquet energy spectrum
\cite{Melin2,Melin3}. Interestingly, these Floquet-Wannier-Stark (FWS)
ladders in the presence of quartets exhibit Landau-Zener-Stückelberg
interference patterns \cite{Melin3,Shevchenko,Dupont-Ferrier}.

In this Article, we present analytical calculations of the FWS ladder
  spectrum in superconducting multiterminal QD, in the limit of small
  dc voltage bias. In this limit, we can use the semiclassical
  approximation, which shows that the FWS spectrum is controlled by
  the value of a Berry phase.  We find that, if the quantum dot level sits at zero energy,
  a {non-trivial} Berry phase $\varphi_B = \pi$ can develop under commensurate voltage
  biasing on the quartet line. We obtain the Bohr-Sommerfeld
  quantization condition by matching the semiclassical wave-functions
  between the different pieces of the classical trajectories in phase
  space. We use the quartet phase and superconducting contact
transparencies as a parameter space which is divided in two regions
with $\varphi_B =0$ and $\varphi_B =\pi$, separated by a hypersurface
  on which the gap closes between the dynamically generated Andreev
  bands.  Finally, we confirm our analytical theory
  by obtaining evidence for the characteristic half-a-period spectral
  shift in the FWS ladder spectrum for $\varphi_B=\pi$, from a
  numerical calculation of the quantum dot tunnel density of states.

This paper is organized as follows. In Sec. II, we introduce the Hamiltonian used as a model of multiterminal superconducting quantum dots.
We develop a tight binding model in Floquet space for these systems in section III.
The adiabatic limit, relevant for small dc voltage biases, is presented in section IV, in 
which the FWS spectrum is shown to depend on a Berry phase. This phase is controlled by a winding number,
whose phase diagram in parameter space is shown.
The tunneling spectra together with the numerical results on the shifted FWS ladder induced
by the non-trivial Berry phase are presented in Sec. V. Summary and perspectives are provided in Sec. VI.
The Appendix gives a detailed presentation of semi-classical calculations, aimed at evaluating the
first non-analytic corrections in $\Delta/V$, which arise from Landau-Zener-St\"uckelberg transitions
between the two FWS ladders originating from the two ABS bands.

\section{Hamiltonian}

We consider in this paper a quantum dot coupled to $N$ superconducting
leads, which are biased at commensurate dc voltages $V_{i}$ ($1 \leq
i \leq N$). We write $V_{i}=s_{i}V$ where $s_{i}$ is an integer (see
Fig.~\ref{fig:0}). For 
example, in the so-called quartet configuration, we have $N=3$, and
$s_{i}=0,\,\pm 1$.

The Hamiltonian of the superconducting-quantum dot takes the following form:
\begin{equation}
H(t)=H_0+H_J(t),\label{eqhmfull}
\end{equation}
where $H_0$ is the BCS Hamiltonian for the superconducting
leads, and $H_J(t)$ describes the tunneling processes between
these reservoirs and the quantum dot.  Specifically:

\begin{widetext}
  \begin{equation}
    H_0 = \sum_{j=1}^{N} \sum_{\sigma} \int \frac{d^{D}\kbf}{(2\pi)^D}
    \left(\epsilon(j,\kbf)c^\dagger_{\sigma}(j,\kbf)c_{\sigma}(j,\kbf) +
    \Delta_j c^\dagger_{\uparrow}(j,\kbf)c^\dagger_{\downarrow}(j,-\kbf)
    +\Delta_{j}^{*}c_{\downarrow}(j,-\kbf)c_{\uparrow}(j,\kbf)
    \right)
\end{equation}
\begin{equation}
  H_J = \sum_{j=1}^{N} J_{j} \sum_{\sigma} \int \frac{d^{D}\kbf}{(2\pi)^D} \left(e^{-is_{j}\omega_{0}t}c^\dagger_{\sigma}(j,\kbf)d_{\sigma}+
  e^{is_{j}\omega_{0}t}d^{\dagger}_{\sigma}c_{\sigma}(j,\kbf)\right)
  \label{eq:H-eq3}
  .
\end{equation}
\end{widetext}
Here $c^\dagger_{\sigma}(j,\kbf)$ and $c_{\sigma}(j,\kbf)$ are
creation and annihilation operators for an electron on reservoir $j$
with momentum $\kbf$ and spin $\sigma$ along the quantization
axis. The corresponding operators on the dot are denoted by
$d^{\dagger}_{\sigma}$ and $d_{\sigma}$. The dimension $D$ of the
reservoirs is denoted by $D$ (with $D=3$ in all numerical
calculations).  The basic frequency $\omega_{0}$ is associated to
single electron tunneling processes, and it is equal to
$\omega_{0}=eV/\hbar$. We have $\omega_{0}=\omega_{J}/2$, where
$\omega_{J}$ is the Josephson frequency associated to $V$. For
simplicity, we assume that the superconducting gaps in all reservoirs
take the same value $\Delta$, and we use the notation
$\Delta_{j}=\Delta e^{i\varphi_{j}}$.

\section{Floquet quasi-particle operators}
\label{section_Floquet_qp_operators}
 
\subsection{Reduction to a 1D chain} 
 
After eliminating the superconducting leads, the Floquet theory of
the time-periodic Bogoliubov-de Gennes equations produces an effective
one-dimensional (1D) tight binding model for the two-component Nambu
spinors $\Psi_m$ describing the part of the wave-function located on
  the dot. Here, we describe the corresponding demonstration of the 1D tight binding model.

The Hamiltonian given by Eqs.~(\ref{eqhmfull})-(\ref{eq:H-eq3}) is
quadratic in the basic fermion operators. Then, the many-body problem
reduces to the simpler time-dependent Bogoliubov-de Gennes equations \be
i\frac{d}{dt}\Gamma^{\dagger}(t)=[H(t),\Gamma^{\dagger}(t)]
\label{general_BdG_equation}
, \ee where $\Gamma^{\dagger}(t)$ denotes a quasi-particle creation
operator. Because the Hamiltonian is periodic in time with period
$T=2\pi/\omega_{0}$, the Floquet theorem leads to \be
\Gamma^{\dagger}(t+T)=e^{-iE T/\hbar}\Gamma^{\dagger}(t).  \ee Taking
the hermitian conjugate of Eq.~(\ref{general_BdG_equation}) leads to
another Floquet solution $\Gamma(t)$, with $E$ changed into its
opposite.

The Fourier series of $\Gamma^{\dagger}_{\sigma}(t)$ is the following:
\begin{widetext}
\be
\Gamma^{\dagger}_{\sigma}(t) =  e^{-iEt/\hbar}\sum_{m \in \mathbb{Z}}
e^{-im\omega_{0}t}\left(u_{m}d^{\dagger}_{\sigma}+\sigma v_{m}d_{-\sigma}+ 
\sum_{i=1}^{N} \int \frac{d^{D}\kbf}{(2\pi)^D}
(u_{m}(i,\kbf)c^{\dagger}_{\sigma}(i,\kbf)+\sigma v_{m}(i,\kbf)c_{-\sigma}(i,\kbf))\right),
\ee
\end{widetext}
where $E$ is the Floquet quasi-energy.
Substituting into Eq.~(\ref{general_BdG_equation}) leads to
\bea
(E+m\omega_{0}-\epsilon(i,\kbf)+i\eta)u_{m}(i,\kbf) & = & \Delta_{i} v_{m}(i,\kbf) + J_{i} u_{m-s_{i}}
\label{u_res} \\
(E+m\omega_{0}+\epsilon(i,\kbf)+i\eta)v_{m}(i,\kbf) & = & \Delta^{*}_{i} u_{m}(i,\kbf) - J_{i} v_{m+s_{i}}
\label{v_res}
,
\eea
and
\bea
(E+m\omega_{0}+i\eta)u_{m} & = & \sum_{i=1}^{N} J_{i}  \int \frac{d^{D}\kbf}{(2\pi)^D} u_{m+s_{i}}(i,\kbf) 
\label{u_dot} \\
(E+m\omega_{0}+i\eta)v_{m} & = & - \sum_{i=1}^{N} J_{i}  \int \frac{d^{D}\kpbf}{(2\pi)^D} v_{m-s_{i}}(i,\kbf)
\label{v_dot}
.  \eea Here, we have introduced a small positive imaginary part
$\eta$ to the quasi-energies. Eliminating the amplitudes in the
reservoirs using Eqs.~(\ref{u_res}), (\ref{v_res}), and substituting
into Eqs.~(\ref{u_dot}), (\ref{v_dot}) leads to

\bea
(E+m\omega_{0}+i\eta)u_{m} -\sum_{i=1}^{N}J_{i}^{2} \; U^{(i)}_{m} & = & 0  \label{L_op_U} \\
(E+m\omega_{0}+i\eta)v_{m} -\sum_{i=1}^{N}J_{i}^{2} \; V^{(i)}_{m} & = & 0 \label{L_op_V}
,
\eea
where
\begin{widetext}
\bea
U^{(i)}_{m} & = & g^{(i)}_{11}(E+(m+s_{i})\omega_{0})u_{m}-
g^{(i)}_{12}(E+(m+s_{i})\omega_{0})v_{m+2s_{i}} \label{def_U_imjk} \\
V^{(i)}_{m} & = & -g^{(i)}_{21}(E+(m-s_{i})\omega_{0})u_{m-2s_{i}}+
g^{(i)}_{22}(E+(m-s_{i})\omega_{0})v_{m} \label{def_V_imjk}
.
\eea
\end{widetext}
Later, we will make extensive use of the linear operator acting on the
collection of amplitudes $u_{m}$, $v_{m}$, which appears on the left
hand side of Eqs.~(\ref{L_op_U}), (\ref{L_op_V}).  This operator will
be denoted by $\mathcal{L}(E)$. We have shown in a previous
work~\cite{Melin3} that all single particle creation and annihilation
operators can be expressed in terms of the resolvent operator
$\mathcal{R}(E)=\mathcal{L}(E)^{-1}$.  The function
$g^{(i)}_{ab}(\omega)$ is the Fourier transform of the retarded
Green's function $g^{(i)}_{\mathrm{ret}}(t)$ of the isolated reservoir
$i$ on the tunneling site connected to the dot, defined as
\be g^{(i)}_{\mathrm{ret}}(t)=-i\left(\begin{array}{cc}
  \{\Psi_{i,\sigma}(t),\Psi^{\dagger}_{i,\sigma}(0)\} & \sigma
  \{\Psi_{i,\sigma}(t),\Psi_{i,-\sigma}(0)\} \\ \sigma
  \{\Psi^{\dagger}_{i,-\sigma}(t),\Psi^{\dagger}_{i,\sigma}(0)\} &
  \{\Psi^{\dagger}_{i,-\sigma}(t),\Psi_{i,-\sigma}(0)\}
\end{array}\right)
\nonumber
\ee
for $t>0$ and $g^{(i)}_{\mathrm{ret}}(t)=0$ for $t<0$.
Here $\Psi_{i\sigma}=\int \frac{d^{D}\kpbf}{(2\pi)^D} c_{\sigma}(i,\kpbf)$.
Explicitely, assuming that $\Im \omega >0$, we have:
\be
g^{(i)}(\omega)=\int \frac{d^{D}\kpbf}{(2\pi)^D\mathcal{D}(\omega,i,\kpbf)}
\left(\begin{array}{cc} \omega+\epsilon(i,\kpbf) & \Delta_{i} \\
  \Delta^{*}_{i} & \omega-\epsilon(i,\kpbf)\end{array}\right)
\nonumber
\ee
where $\mathcal{D}(\omega,i,\kpbf)=\omega^{2}-\epsilon(i,\kpbf)^{2}-|\Delta_{i}|^{2}$.

Let us introduce the family of two-component spinors
$\Psi_{m}=(u_m,v_m)^{T}$, labeled by $m$.  We focus on the case of
three reservoirs ($N=3$), with dc bias voltages in the quartet
configuration: $s_{a}=-1$, $s_{b}=1$, and $s_{c}=0$. In this case,
the homogeneous Eqs.~(\ref{L_op_U}) and (\ref{L_op_V}) take the
form \be
M_{0}(m)\Psi_{m}-M_{+}(m+1)\Psi_{m+2}-M_{-}(m-1)\Psi_{m-2}=0,
\label{homogeneous_difference_eq}
\ee

The off-diagonal terms 
    in $m$ are second order Andreev reflection processes between the dot and the reservoirs, which explains  
    why $m$ is coupled to $m\pm 2$.
The expanded forms of the matrices $M_{0}(m)$ and $M_{\pm}(m)$ are
presented in the following subsection.  

\subsection{Explicit forms of $M_{0}(m)$ and $M_{\pm}(m)$}
\label{subsec_explicit_forms}
Now, we provide the expression of the the matrices $M_0$ and $M_\pm$
[see Eq.~(\ref{homogeneous_difference_eq})]. To simplify the discussion, we
assume a constant density of states $\rho_{0}$ in the normal state. We
take the Fermi energy at $\epsilon_F=0$, and assume an infinite
bandwidth, which implies exact particle-hole symmetry in the leads.
This suggests to introduce the integral \be
\label{eq:gh1}
I(E)=\rho_{0}\int_{-\infty}^{\infty}\frac{d\epsilon}{E^{2}-|\Delta|^{2}-\epsilon^{2}}
.
\ee Here, we are interested in the retarded Green's function, and
an infinitesimal positive imaginary part is added to
energy $E$. Then, Eq.~(\ref{eq:gh1}) takes the form
\bea I(E) & = & \frac{-\pi\rho_{0}}{\sqrt{|\Delta|^{2}-E^{2}}},\;\;
E^{2}<|\Delta|^{2} \\ I(E) & = & \frac{-i
  \pi\rho_{0}}{\sqrt{E^{2}-|\Delta|^{2}}}\:\mathrm{sign}(E),\;\;
E^{2}>|\Delta|^{2}. \eea
The retarded Green's function is then given by \be
g(\omega)=I(\omega) \left(\begin{array}{cc} \omega & \Delta
  \\ \Delta^{*} & \omega \end{array}\right) \nonumber .\ee
\begin{widetext}
 
  Let us now give explicit expressions for the $M_{0}(m)$ and
  $M_{\pm}(m)$ matrices introduced in Eq.~(\ref{homogeneous_difference_eq}). The matrices
  depend also on the energy $E$. We
  introduce the variable $\xi=m\omega_{0}$, where
  $\omega_{0}=eV/\hbar$.  The density of states in reservoir $j$ is
  denoted by $\rho_{0,j}$. It is also convenient to define
  $\Gamma_{j}=\pi \rho_{0,j}J_{j}^{2}$.  We assume $\Delta_{j}=\Delta
  e^{i\varphi_{j}}$. Global gauge invariance allows us to set
  $\varphi_{c}=0$.  In the case $|E+\xi|<\Delta$, we have: \be
  M_{0}(m)=\left(\begin{array}{cc}
    (E+\xi)\left(1+\frac{\sum_{j}\Gamma_{j}}{\sqrt{\Delta^{2}-(E+\xi)^{2})}}\right)
    &
    -\frac{\Gamma_{c}\Delta}{\sqrt{\Delta^{2}-(E+\xi)^{2})}}\\ -\frac{\Gamma_{c}\Delta}{\sqrt{\Delta^{2}-(E+\xi)^{2})}}
    &
    (E+\xi)\left(1+\frac{\sum_{j}\Gamma_{j}}{\sqrt{\Delta^{2}-(E+\xi)^{2})}}\right)
  \end{array}\right)
\ee
\be
M_{+}(m)=\left(\begin{array}{cc} 0 &  \frac{\Gamma_{b}\Delta e^{i\varphi_{b}}}{\sqrt{\Delta^{2}-(E+\xi)^{2})}} \\
  \frac{\Gamma_{a}\Delta e^{-i\varphi_{a}}}{\sqrt{\Delta^{2}-(E+\xi)^{2})}} & 0
  \end{array}\right),\;\;\;\;\;\mbox{and}\;\;\;\;\;
M_{-}(m)=\left(\begin{array}{cc} 0 & \frac{\Gamma_{a}\Delta e^{i\varphi_{a}}}{\sqrt{\Delta^{2}-(E+\xi)^{2})}} \\
  \frac{\Gamma_{b}\Delta e^{-i\varphi_{b}}}{\sqrt{\Delta^{2}-(E+\xi)^{2})}} & 0
\end{array}\right)
.
\ee

In the case $|E+\xi|>\Delta$, these expressions become
\be
M_{0}(m)=\left(\begin{array}{cc} (E+\xi)\left(1+i\frac{\sum_{j}\Gamma_{j}}{\sqrt{(E+\xi)^{2}-\Delta^{2}}}\right) &
  -\frac{i\Gamma_{c}\Delta}{\sqrt{(E+\xi)^{2}-\Delta^{2}}}\\
  -\frac{i\Gamma_{c}\Delta}{\sqrt{(E+\xi)^{2}-\Delta^{2}}}
  & (E+\xi)\left(1+i\frac{\sum_{j}\Gamma_{j}}{\sqrt{(E+\xi)^{2}-\Delta^{2}}}\right)
  \end{array}\right)
\label{general_M_0_above_gap}
\ee
\be
M_{+}(m)=\left(\begin{array}{cc} 0 &  \frac{i\Gamma_{b}\Delta e^{i\varphi_{b}}}{\sqrt{(E+\xi)^{2}-\Delta^{2}}} \\
  \frac{i\Gamma_{a}\Delta e^{-i\varphi_{a}}}{\sqrt{(E+\xi)^{2}-\Delta^{2}}} & 0
\end{array}\right),
\label{general_M_+-_above_gap}
\;\;\;\;\;\mbox{and}\;\;\;\;\;
M_{-}(m)=\left(\begin{array}{cc} 0 & \frac{i\Gamma_{a}\Delta e^{i\varphi_{a}}}{\sqrt{(E+\xi)^{2}-\Delta^{2}}} \\
  \frac{i\Gamma_{b}\Delta e^{-i\varphi_{b}}}{\sqrt{(E+\xi)^{2}-\Delta^{2}}} & 0
\end{array}\right)
\ee
From these matrices, we build the $2\times 2$ matrix $L_{0}(\xi,k)$. This matrix will be used to obtain the classical trajectories according to $\mathrm{det}\;L_{0}(\xi,k)=0$ in our semiclassical treatment in the forthcoming section. We have:
\bea
L_{0}(\xi,k) & = & \left(\begin{array}{cc} (E+\xi)\left(1+\frac{\sum_{j}\Gamma_{j}}{\sqrt{\Delta^{2}-(E+\xi)^{2})}}\right) &
  -\frac{(\Gamma_{a}e^{i(\varphi_{a}-k)}+\Gamma_{b}e^{i(\varphi_{b}+k)}+\Gamma_{c})\Delta}{\sqrt{\Delta^{2}-(E+\xi)^{2})}}\\
  -\frac{(\Gamma_{a}e^{-i(\varphi_{a}-k)}+\Gamma_{b}e^{-i(\varphi_{b}+k)}+\Gamma_{c})\Delta}{\sqrt{\Delta^{2}-(E+\xi)^{2})}}
  & (E+\xi)\left(1+\frac{\sum_{j}\Gamma_{j}}{\sqrt{\Delta^{2}-(E+\xi)^{2})}}\right)
\end{array}\right),\;\;\;\;\;\mbox{ if $|E+\xi|<\Delta$}
\label{explicit_L0_inside_gap} \\
L_{0}(\xi,k) & = & \left(\begin{array}{cc} (E+\xi)\left(1+\frac{i\sum_{j}\Gamma_{j}}{\sqrt{(E+\xi)^{2}-\Delta^{2}}}\right) &
  -\frac{i(\Gamma_{a}e^{i(\varphi_{a}-k)}+\Gamma_{b}e^{i(\varphi_{b}+k)}+\Gamma_{c})\Delta}{\sqrt{(E+\xi)^{2}-\Delta^{2}}}\\
  -\frac{i(\Gamma_{a}e^{-i(\varphi_{a}-k)}+\Gamma_{b}e^{-i(\varphi_{b}+k)}+\Gamma_{c})\Delta}{\sqrt{(E+\xi)^{2}-\Delta^{2}}}
  & (E+\xi)\left(1+\frac{i\sum_{j}\Gamma_{j}}{\sqrt{(E+\xi)^{2}-\Delta^{2}}}\right)
\end{array}\right),\;\;\;\;\;\mbox{ if $|E+\xi|>\Delta$}
\label{explicit_L0_above_gap}
.
\eea
\end{widetext}

They explicitly depend on
the quasiparticle Floquet energy $E$,  but only {\it via } the
combination $E+m\omega_{0}$, where
$\omega_{0}=eV/\hbar$. This allows us to interpret Eq.~(\ref{homogeneous_difference_eq})
as the Schr\"odinger equation for a 1D Floquet
tight-binding Hamiltonian which
contains a fictitious
uniform
  electric field $\omega_0$, related to the energy $-m\omega_0$ of the
  Cooper pairs transmitted by Andreev reflection
  in the superconducting leads. For such tight-binding models
\cite{Wannier,Bentosela}, the energy spectrum consists of several
Wannier-Stark ladders, each containing equally
spaced levels separated by $\hbar\omega_0$. In
addition, for the superconducting-QD of interest,
the Floquet states are connected by multiple Andreev reflections to
the superconducting quasiparticle continua in the leads, if
$|E+m\omega_{0}|> \Delta$ (with $\Delta$ the superconducting
gap). This provides a finite life-time (or equivalently a finite
spectral width) to the FWS resonances \cite{Melin2,Melin3}.

\section{Adiabatic approximation}

\subsection{Zero voltage limit}

In a three-terminal superconducting-QD, the condition {for emergence
  of quartets is set by commensurate voltage biasing
  $(V_a\,,V_b,\,V_c)=(V,\,-V,\,0)$ on the superconducting leads
  $S_a,\,S_b$ and $S_c$ \cite{Freyn1}.}  {The matrices $M_{0}(m)$ and
  $M_{\pm}(m)$ no longer depend on $m$ in the ``classical'' limit
  $V=0$.  We can then use Bloch theorem to solve
  Eq.~(\ref{homogeneous_difference_eq}), which produces} plane-wave
solutions $\Psi_{m}=\exp\left(ikm/2\right)\Psi$. {The wave vector $k$
  appears as a free parameter and it can be physically interpreted by
  noting that the adiabatic approximation for the time-dependent
  problem becomes exact if $V \rightarrow 0$. These} plane-wave
solutions correspond to the quasiparticle operators for {{\it static}
  Bogoliubov-De Gennes Hamiltonians} with {the superconducting
  order-parameter} phases {given by} \be
\varphi_{j}(k)=\varphi_{j}+s_{j}k
\label{phases_versus_k}
, \ee {where $s_j=\pm 1,\,0$ according to the voltage $V_j=\pm V,\,0$
  on lead $S_j$.} The doublet of ABS bands has then energy dispersion
relation $E=\pm E_A(k)$, which is {a $2\pi$-periodic} function of {the
  analogous wave-vector} $k$.  The first task here is to calculate this
dispersion relation, including self-energy corrections due to the
reservoirs. It is easy to show that it is determined by solving the equation:
\be
\mathrm{det}\left(L_{0}(\xi=0,k)\right),
\label{adiabatic_eigenvalue_eq}
\ee
where $E=\pm E_A(k)$ lies inside the superconducting gap, so Eq.~(\ref{explicit_L0_inside_gap})
has to be used to define the two by two matrix $L_{0}$.

\subsection{Andreev bound-state dispersion relation}
\label{subsec:ABS_dispersion}

In our three terminal setting biased in the quartet configuration,
Eq.~(\ref{adiabatic_eigenvalue_eq}) takes the form
\be
f(x)=\pm\frac{|\Gamma(k)|}{\Delta}
\label{explicit_eq_classical_trajectory_2}
\ee with $x=E_{A}(k)/\Delta$, $f(x)=x(\sqrt{1-x^{2}}+c)$,
$c=\sum_{j}\Gamma_{j}/\Delta$, and
$\Gamma(k)=\Gamma_{a}e^{i(\varphi_{a}-k)}+\Gamma_{b}e^{i(\varphi_{b}+k)}+\Gamma_{c}$, 
using a gauge in which $\varphi_c=0$. This provides an implicit determination
of $E_{A}(k)$. Since this 
equation is valid inside the BCS gap in the reservoirs, it requires that
$|x|<1$.  When $x$ increases from $0$, $f(x)$ first increases, it
reaches a maximum at $x=x_{M}$, and then decreases until $x=1$.
Explicitely, $x_{M}=\sqrt{4-c^{2}+c\sqrt{8+c^{2}}}/\sqrt{8}$. In the
tunnel limit, $c\ll 1$, $x_{M}\simeq \sqrt{2}/2$. We have the useful
inequality: \be 0<\frac{|\Gamma(k)|}{\Delta}\leq c=f(1). \ee Note that
$|\Gamma(k)|/\Delta=c$ only if
$\exp({i(\varphi_{a}-k)})=1=\exp({i(\varphi_{b}+k)})$, which implies that
$\varphi_{q}=\varphi_{a}+\varphi_{b}=0\;\mathrm{mod}\;2\pi$.  When
$\varphi_{q} \neq 0\;\mathrm{mod}\;2\pi$, for any $k$, there is a
unique solution to Eq.~(\ref{explicit_eq_classical_trajectory_2}) with
$0<x(k)<x_{M}$. Some examples of ABS dispersion relations are shown as the magenta
curves on panels (b) and (d) of Fig.~\ref{fig:semi_class_trajectory} in the Appendix.

In the tunnel limit, when $\Gamma_{j}\ll \Delta$, solutions of
Eq.~(\ref{explicit_eq_classical_trajectory_2}) satisfy $|x(k)|\ll 1$,
and $f(x)$ can be well approximated by its tangent near the origin,
i.e. $f(x)\simeq cx$. This approximation amounts to neglecting the
energy dependence of self energy corrections, at least in the subgap
region, and we will use it quite often in the following
discussions. This corresponds to making the following approximation:
\be L_{0}(\xi=0,k)\simeq\left(\begin{array}{cc} (1+c)E &
  -\Gamma(k)\\ -\Gamma(k)^{*} & (1+c)E
  \end{array}\right)
\label{L_0_simplified}
\ee
In this case, Eq.~(\ref{explicit_eq_classical_trajectory_2}) becomes:
\be
E_{A}(k)=\pm\frac{|\Gamma(k)|}{1+c}
\label{E_ABS_tunnel_limit}
\ee

\begin{figure}[htb]
    \includegraphics[width=\columnwidth]{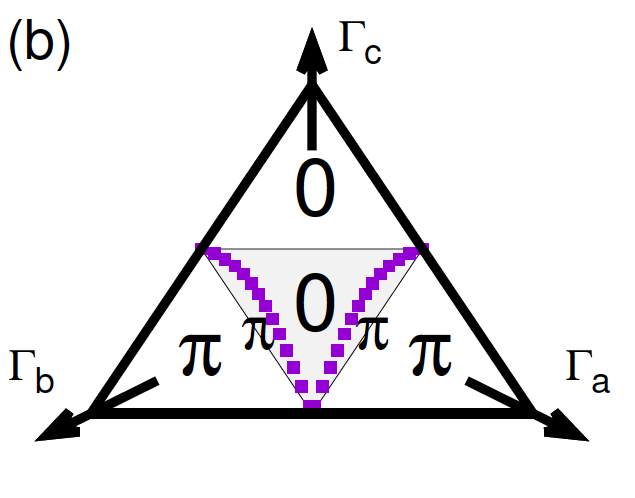}
    \caption{{\it Ternary diagram for the gap
        closing condition:} The nodal lines, displayed in magenta and
      calculated for $\varphi_q/2\pi = 0.2$, represent the values of
      the parameters for which the gap between the two Andreev bound
      state bands vanishes [see Eq.~(\ref{eq:Gammac-0})]; below these
      two lines, the Berry {phase takes the value $\varphi_B = \pi$.}
      The smaller shaded inner triangle shows all the possible values
      of the nodal lines when $0<\varphi_q/2\pi<1$.
      \label{fig.Berry_phase_diagram}
    }
\end{figure}

The gap between the two Andreev bound-state bands closes when there
is at least one value of $k$ such that $x(k)=0$, which requires
$\Gamma(k)=0$. For this to happen, the triangular inequality
$|\Gamma_{a}-\Gamma_{b}| \leq \Gamma_{c} \leq \Gamma_{a}+\Gamma_{b}$
has to be satisfied [see the shaded inner triangle on Fig.~\ref{fig.Berry_phase_diagram}].
If this is the case, there are two angles $\alpha$ and
$\beta$, lying in $]-\pi,\pi[$, whose values depend on $\Gamma_{j}$'s,
such that $x(k)=0$ if and only if $(\varphi_{a}-k,\varphi_{b}+k)=\pm(\alpha,\beta)$.
This shows that, generically (precisely when $\Gamma_{a} \neq \Gamma_{b}$), the gap closes for two different values of
$\varphi_{q}=\pm(\alpha+\beta)$.  For each of them, there is a
unique value of $k$ such that $x(k)=0$. The gap closes at
$\varphi_q=0$ mod. $2\pi$ if $\Gamma_a=\Gamma_b$ and
$\alpha+\beta=0$ mod. $2\pi$, and there are two values of $k$ such
that $x(k)=0$. This gap closing condition can be formulated as
follows in the generic case $\Gamma_{a} \neq \Gamma_{b}$ or
$\varphi_{q} \neq 0$ mod. $2\pi$: the gap closes if $\Gamma_c=\Gamma_c^{(0)}$, with
\be
\Gamma_{c}^{(0)}=\frac{|\Gamma_{a}^{2}-\Gamma_{b}^{2}|}{\sqrt{\Gamma_{a}^{2}+\Gamma_{b}^{2}-2\Gamma_{a}\Gamma_{b}\cos\varphi_{q}}}.
\label{eq:Gammac-0}
\ee
This relation is represented by the magenta datapoints on Fig.~\ref{fig.Berry_phase_diagram}.

\subsection{Floquet energies in the adiabatic limit}

In the adiabatic limit, the solution of the time-dependent Bogoliubov-de Gennes
Eq.~(\ref{general_BdG_equation}) is well approximated by:
\be
\Gamma^{\dagger}(t)=e^{-i\varphi(t)}\Gamma^{\dagger}_{A}(t), 
\ee
where $\Gamma^{\dagger}_{A}(t)$ is a quasiparticle creation operator associated to the Andreev
bound-state for the Hamiltonian $H(t)$. This means that $\Gamma^{\dagger}_{A}(t)$
satisfies:
\be
[H(t),\Gamma^{\dagger}_{A}(t)]=E_{A}(t)\Gamma^{\dagger}_{A}(t).
\ee
The physical time variable $t$ is directly related to the the wave-vector $k$ discussed
earlier by $k=2\omega_{0}\:t=\omega_{J}\:t$.
As usual~\cite{Berry2}, the phase-factor $\varphi(t)$ is the sum of two contributions,
a dynamical phase $\varphi_{d}$ and a geometrical phase $\varphi_{g}$. As always:
\be
\varphi_{d}=\frac{1}{\hbar}\int_{0}^{t}E_{A}(t')\:dt'.
\label{dynamical_phase}
\ee
The geometrical phase $\varphi_{g}$ depends generically on an arbitrary choice of phases
for the instantaneous quasi-particle operators $\Gamma^{\dagger}_{A}(t)$, excepted when the system
Hamiltonian at time $t$ is the same as at $t=0$, in particular when $t$ is equal to the Josephson
period $T_{J}=2\pi/\omega_{J}$. In this case, $\varphi_{g}$ is gauge invariant, and is an example of
a Berry phase $\varphi_B$.

To evaluate this Berry phase, we have to describe in more detail how the
quasi-particle operators $\Gamma^{\dagger}_{A}(k)$ vary as $k$ evolves from $-\pi$ to $\pi$.
Writing:
\be
\Gamma^{\dagger}_{A\sigma}(k)=u(k) d^{\dagger}_{\sigma}+\sigma v(k) d_{-\sigma}+...,
\label{explicit_Gamma_A}
\ee
where the dots refer to virtual contributions from the superconducting reservoirs,
and the spinor  $\chi(k)=(u(k),v(k))^{T}$ is a nul eigenvector for $L_{0}(\xi=0,k)$, i.e.
$L_{0}(\xi=0,k)\chi(k)=0$, with $E=E_{A}(k)$ in $L_{0}(\xi=0,k)$. From Eq.~(\ref{L_0_simplified}), we see
that the Nambu spinor $\chi(k)$ associated to the dot is subjected to a fictitious magnetic field
lying in the $x-y$ plane and oriented along $\Gamma(k)$ (after identifying complex numbers
with points in the $x-y$ plane in the usual way).
As $k$ runs from $-\pi$ to $\pi$, $\Gamma(k)$ describes an ellipse $\mathcal{E}$ around
the origin of the complex plane.  For such closed path, the winding number $w$ is defined as
\be \alpha(\pi)-\alpha(-\pi)=2\pi w,
\ee
where $\Gamma(k)=|\Gamma(k)|e^{i\alpha(k)}$.
Correspondingly, the pseudo-spin associated to the Nambu spinor $\chi(k)$
  performs $w$ turns around the equator on the Bloch sphere, which
  induces a Berry phase $\varphi_B\equiv -w\pi$~\cite{Berry2}. The
  appearance of Berry phases in multi-component WKB equations has been
  pointed out by many authors, both in the
  mathematics~\cite{Guillemin} and in the
  physics~\cite{Wilkinson,Wang} communities. In this specific model,
  when the dot level lies exactly at zero energy,
  $\varphi_B$ takes only two values: 0 or $\pi$ modulo $2\pi$.

  At this point, we should emphasize that this quantization is not robust. A finite gate
  voltage acting on the the dot adds a term proportional to the Pauli matrix $\sigma^{z}$
  in $L_{0}(\xi=0,k)$. As a result, the spinor $\chi(k)$ is no longer confined to the equator, but
  to a constant altitude circle on the Bloch sphere. In this situation, the Berry phase is now equal
  to $w \zeta$, where $\zeta=\pi$ at zero gate voltage, and departs from $\pi$ linearly at small gate voltage.

We have $w=0$ so $\varphi_B=0$ when the
origin of the complex plane is not inside the ellipse $\mathcal{E}$,
and $w=\pm 1$ so $\varphi_B= \pm \zeta$ otherwise. This implies that $\varphi_B$ jumps
from $0$ to $\pm \zeta$, precisely at the point in parameter space where the
minimum over $k$ of $|\Gamma(k)|$ vanishes.
Interestingly, we always have $\varphi_B\equiv \pm \zeta$ in the two-terminal case,
so that the third terminal biased is necessary
in order to observe the jump of the Berry phase from $\pm \zeta$ to $0$.

A cautious reader may worry that the Berry phase might be sensitive to 
virtual contributions from the superconducting reservoirs, whose existence is reminded
by the dots in Eq.~(\ref{explicit_Gamma_A}). In fact, such virtual contributions are
fully taken into account through the self-energies, which appear in the expression~(\ref{explicit_L0_inside_gap}) for $L_{0}(\xi,k)$.
So the previous discussion of the Berry phase does take into account these virtual contributions.

The Berry phase $\varphi_{B}$ manifests itself on the Floquet spectrum, as we request
that the quasiparticle operator $\Gamma^{\dagger}(t)$ should satisfy the periodicity condition
\be
\Gamma^{\dagger}(t+T_{J})=e^{-iET_{J}}\Gamma^{\dagger}(t).
\ee
This leads to
\be
\varphi_{d}(T_{J})+\varphi_{B}=ET_{J}+2\pi n,
\ee
where $n$ is an arbitrary integer. Because $\varphi_{d}(T_{J})=\langle E_{A} \rangle T_{J}$,
where $\langle E_{A} \rangle$ denotes the Andreev bound-state energy averaged over one period,
this leads to the following Bohr-Sommerfeld type formula:
\be
E=\langle E_{A} \rangle-(2n+w \frac{\zeta}{\pi})\omega_{0}
\label{adiabatic_Floquet_spectrum}
\ee
Because of the charge conjugation symmetry of the Bogoliubov-de Gennes equations, applying
hermitian conjugation to $\Gamma^{\dagger}(t)$ produces a second Wannier-Stark ladder in which
$E$ is replaced by $-E$ modulo $\omega_{J}$. At the level of the adiabatic approximation presented
in this subsection, these two Wannier-Stark ladders remain decoupled. 

\subsection{Floquet spectrum beyond the adiabatic limit}

Going beyond the adiabatic approximation does induce some coupling, and physically, this coupling
corresponds to Landau-Zener transitions between the two Andreev bound-state bands. A natural way to
capture them is to go back to the general Floquet formulation presented in section~\ref{section_Floquet_qp_operators}.
At small dc voltage bias, the linear potential term $m\omega_{0}$ entering in Eqs.~(\ref{u_res}) to (\ref{v_dot}) is small,
so the difference equation~(\ref{homogeneous_difference_eq}) can be treated by the WKB method.
Since this is a standard method, and given that explicit calculations are a bit tedious, we have relegated this contribution to
the Appendix. Interestingly, the shift in the Floquet spectra induced by a non trivial Berry phase is still
present at intermediate voltages (compared to the superconducting gap). To cover the full range of possible
voltages, a full numerical calculation is necessary, whose results will be presented on
Fig.~\ref{fig:spectroscopic_manifestation_W} below.

\section{Tunneling spectra calculations}

Here, we would like to illustrate our prediction by exploring the FWS ladder spectra under the presence of the non-trivial Berry phase.
We evaluate the spectra which could possibly be measured in a superconducting multiterminal quantum dot with an additional tunneling probe,
as depicted in Fig.~\ref{fig:0}.
\begin{figure*}[htb]
   
    \includegraphics[width=\textwidth]{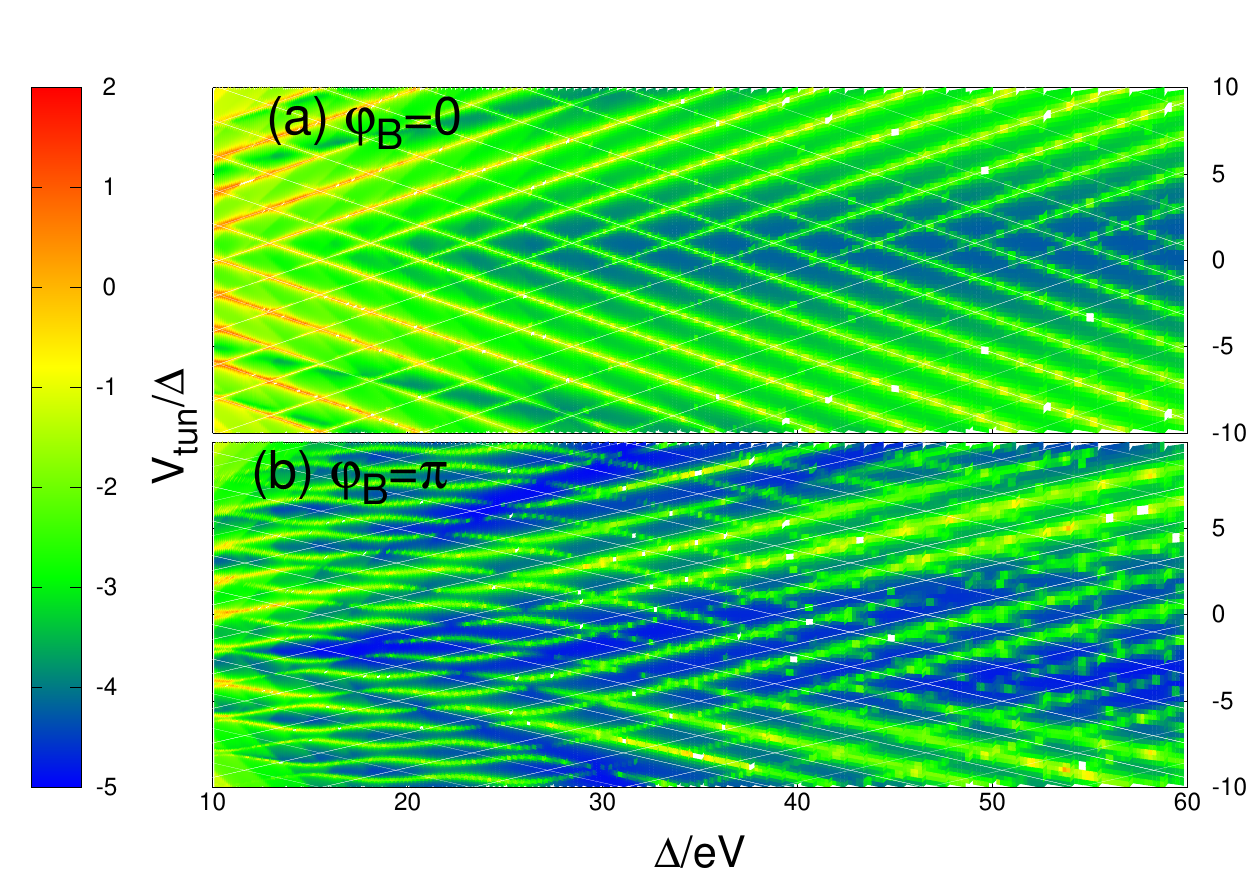}
    \caption{{\it Tunnel spectroscopy of the Berry
          phase:} The figure features the logarithm of the local
        density of states on the quantum dot (in colorscale) as a
        function of inverse voltage $\Delta/eV$ ($x$-axis) and tunnel
        probe bias voltage $eV_{tun}/\Delta$ ($y$-axis). The Berry phase is
        $\varphi_B=0$ on panel (a), and $\varphi_B=\pi$ on panel
        (b). The tunnel spectra reveal the Floquet-Wannier-Stark
        ladders, and they are compared to the tilted white lines which
        correspond to $w=0$ in Eq.~(\ref{simple_WS_ladder}). The
        half-a-period shift appearing on panel (b) is signature of
        nontrivial Berry phase $\varphi_B=\pi$, while $\varphi_B=0$
        for the nonshifted tunnel spectrum on panel (a). The
        three-terminal superconducting-QD has $\Gamma_a/\Delta=0.4$,
        $\Gamma_b/\Delta=0.2$ and $\varphi_q=0$, with (a)
        $\Gamma_c/\Delta=1.0$ and (b) $\Gamma_c/\Delta=0.3$. 
      \label{fig:spectroscopic_manifestation_W}
    }
\end{figure*}

\subsection{Tunneling density of states}
\label{subsec_tunneling}

One way to detect FWS ladders is to perform local tunneling
spectroscopy on the dot. For this, we tunnel-couple a normal probe to
the quantum dot. The differential tunneling conductance is given by:
\begin{widetext}
\be
\frac{\partial I_{\mathrm{tun}}(t)}{\partial V_{\mathrm{tun}}} =-e^{2}\int_{-\infty}^{\infty}\frac{d\omega}{2\pi}J_{\mathrm{tun}}^{2}(\omega)f'_{FD}(\omega+eV_{\mathrm{tun}})
\sum_{n}i\left(G^{R}_{d}(\omega)_{n}-G^{R}_{d}(\omega)_{-n}^{*}\right)e^{-in\omega_{0}t}.
\ee
\end{widetext}
Here, $J_{\mathrm{tun}}^{2}(\omega)$ is the Fermi golden rule squared tunneling amplitude times the density of states in the normal probe at
energy $\omega$, and $f'_{FD}(\omega)$ is the derivative of the Fermi-Dirac distribution. Because of the periodic time dependence 
of the BCS Hamiltonian, the tunneling current $I_{\mathrm{tun}}(t)$ is also periodic in time. 
$G^{R}_{d}(\omega)_{n}$ is the Fourier transform of the retarded Green's function on the dot, defined explicitely as:
\be
G^{R}_{d}(t,t')=\int_{-\infty}^{\infty}\frac{d\omega}{2\pi} \sum_{n} G^{R}_{d}(\omega)_{n} e^{-i\omega (t-t')}e^{-in\omega_{0}t}.
\ee
In fact, $G^{R}_{d}(\omega)_{n}$ is directly related to the resolvent operator defined earlier through
$G^{R}_{d}(\omega)_{n}=\mathcal{R}(\omega)_{n,0}$.
The dc tunneling current takes a particularly simple form:
\be
\frac{\partial I_{\mathrm{tun}}}{\partial V_{\mathrm{tun}}} =2e^{2}\int_{-\infty}^{\infty}\frac{d\omega}{2\pi}J_{\mathrm{tun}}^{2}(\omega)f'_{FD}(\omega+eV_{\mathrm{tun}}) \Im G^{R}_{d}(\omega)_{n=0}.
\ee
Finally, in the zero temperature limit, we obtain:
\be
\frac{\partial I_{\mathrm{tun}}}{\partial V_{\mathrm{tun}}} =-2e^{2}J_{\mathrm{tun}}^{2}(-eV_{\mathrm{tun}}) \Im G^{R}_{d}(\omega)_{n=0}.
\ee

\subsection{Numerical results}

These results raise the question of the possible experimental
observation of these effects.  Recently, we have shown that finite
frequency noise measurements provide an experimental access to
differences $E_{n}-E_{n'}$ between two FWS quasi-energy eigenvalues
~\cite{Melin3}.  This is interesting to evidence level repulsion
induced by Landau-Zener-Stückelberg inter ladder tunneling processes,
but this noise spectroscopy is not sensitive to the
global shift of the FWS spectrum induced by Berry phase
jumps. Therefore, we propose to perform
tunnel spectroscopy on the quantum
  dot [see Fig.~\ref{fig:0}(a)]. The differential dc-tunnel
  conductance through the dot directly probes the FWS ladder density
  of states. Fig.~\ref{fig:spectroscopic_manifestation_W} shows two
  tunnel spectra, one for $\varphi_B=0$ [panel (a)], and the other one
  for $\varphi_B=\pi$ [panel (b)]. The global shift associated to a
  Berry phase jump is clearly visible in
  Fig.~\ref{fig:spectroscopic_manifestation_W} while comparing in both
  cases the numerical tunnel spectra to the tilted reference white
  line corresponding to $w=0$ in Eq.~(\ref{simple_WS_ladder}). For further details on the numerical calculation of the resolvent, see Section III in the Supplemental Material \cite{supplemental}.

\section{Summary and perspectives}

To conclude, we have shown that, in superconducting multiterminal
QD, a non-trivial Berry phase $\varphi_B$ can appear on the quartet
line at commensurate voltages. Via semiclassical calculations, we have demonstrated that the parameter
space splits into two regions with $\varphi_B=0$ or $\varphi_B=\pi$,
separated by a hypersurface on which the gap between the Andreev bands
closes. We have seen that the FWS spectrum is controlled by the Berry phase. 
The non-trivial Berry phase can be revealed
  by probing the density of states of the quantum dot in a tunneling spectroscopy experiment. Our numerical calculations directly show that
  the FWS ladder spectra is shifted by half-a-period when $\varphi_B=\pi$, as
  compared to $\varphi_B=0$. While our calculations are performed when the superconducting quantum dot level sits at zero energy, one may expect to continuously tune the Berry phase by changing the energy of the dot, for example via electrostatic gating.

\acknowledgments{The authors thank the Centre R\'egional Informatique
  et d'Applications Num\'eriques de Normandie (CRIANN)
  for use of its facilities.  The authors thank
  the Infrastructure de Calcul Intensif et de Donn\'ees (GRICAD) for
  use of the resources of the M\'esocentre de Calcul Intensif de
  l'Universit\'e Grenoble-Alpes (CIMENT). B.D. and R.M. acknowledge
  fruitful discussions with Y. Colin de Verdi\`ere, {F. Faure and
    A. Joye. R.D. and R.M.} acknowledge financial support from the
  Centre National de la Recherche Scientifique (CNRS) and Karlsruhe
  Institute of Technology (KIT), through the International Laboratory
  ``LIA SUPRADEVMAT'' between the Grenoble and Karlsruhe
  campuses. This work was partly supported by Helmholtz society
  through program STN and the DFG via the projects DA 1280/3-1.}

\appendix

\section{Local semi-classical solutions}

\subsection{General idea}
\label{sub_gen_idea_semi_class}

A small bias voltage $V$ plays formally the role
of Planck's constant $\hbar$ in the Wentzel–Kramers–Brillouin (WKB)
approximation \cite{Landau}. The classical limit 
  $\hbar \rightarrow 0$ in standard quantum mechanics corresponds to
  $eV/\Delta \rightarrow 0$ in superconducting-QD. The semiclassical
  approximation for $eV \ll \Delta$ in superconducting junctions was
pioneered by Bratus' \textit{et al.} \cite{Bratus}. In this
approximation, the wave-vector $k$ has slow
  variations with $m$. Let us first transform $m$ into a continuous variable {\it via}
\begin{equation}
\epsilon=2\omega_0,\quad m\omega_0=\xi,
\end{equation}
where $\epsilon$ is a small parameter. Eq.~(\ref{homogeneous_difference_eq}) reads
\begin{equation}
M_0(\xi)\Psi(\xi)-M_+(\xi+\frac{\epsilon}{2})\Psi(\xi+\epsilon)-M_-(\xi-\frac{\epsilon}{2})\Psi(\xi-\epsilon)=0.
\label{eqovmc}
\end{equation}
The semiclassical Ansatz takes the form \be
\Psi(\xi)=e^{i\frac{\theta(\xi)}{\epsilon}}\chi(\xi), \ee where
$\chi(\xi)$ can be expanded in $\epsilon$ according to \be
\chi(\xi)=\sum_{n=0}^{\infty}\epsilon^{n}\chi_{n}(\xi) .\ee Assuming
that $\theta(\xi)$ and $\chi(\xi)$ have infinitely many derivatives,
we can view the linear operator acting on $\chi(\xi)$
in Eq.~(\ref{eqovmc}) as a differential operator $L$ of infinite order,
which can also be expanded in $\epsilon$ according to \be
L=\sum_{n=0}^{\infty}\epsilon^{n}L_{n}.\ee This leads to an infinite
set of equations, from which we keep the first two of lowest order:
\be L_{0}\chi_{0}(\xi) = 0
\label{order_0_eq}
\ee
\be
L_{0}\chi_{1}(\xi)+L_{1}\chi_{0}(\xi) = 0
\label{order_1_eq}
.
\ee

\subsection{Classical phase-space trajectories}
\label{sub_semi_classical_trajectory}

Let us first consider the
zero-th order Eq.~(\ref{order_0_eq}): \be
L_{0}(\xi,\theta'(\xi))\chi_{0}(\xi)=0,
\label{order_0_eq_2}
\ee where \be
L_{0}(\xi,\theta'(\xi))=M_{0}(\xi)-M_{+}(\xi)e^{i\theta'(\xi)}-M_{-}(\xi)e^{-i\theta'(\xi)}
\ee is a $2\times 2$ matrix. The $L_{0}$ operator acts on
$\chi_{0}(\xi)$ only through point-wise multiplication, {\it i.e.} it
does not involve any differential operator involving the $\xi$
variable.  Eq.~(\ref{order_0_eq}) has nontrivial solutions if
$\mathrm{det}\;L_{0}(\xi,k)=0$, namely \be
\mathrm{det}\left(M_{0}(\xi)-M_{+}(\xi)e^{i\theta'(\xi)}-M_{-}(\xi)e^{-i\theta'(\xi)}\right)=0
\label{general_eq_classical_trajectory_2}
.  \ee
In the general spirit of semiclassical (or WKB) approximation,
we introduce the $\xi$-dependent wave number $k(\xi)$ by
$k(\xi)=\theta'(\xi)$.  Eq.~(\ref{general_eq_classical_trajectory_2})
determines a curve in the $(\xi,k)$ plane called the {\em classical
  trajectory} in {\em phase space}. Recalling that, in
Eq.~(\ref{homogeneous_difference_eq}), $E$ and $m$ enter only through
the combination $E+m\omega_{0}$, we see that $E$ and $\xi$ enter
Eq.(\ref{general_eq_classical_trajectory_2}) only through $E+\xi$. As
a result, this classical trajectory is related in a very simple manner
to the Andreev subbands energy dispersion relation $E_{A}(k)$ by: \be
E+\xi=\sigma E_{A}(k)
\label{key_tilted_band_relation}
\ee where $\sigma=\pm 1$ labels the two Andreev subbands. In
Eq.~(\ref{key_tilted_band_relation}), the total energy $E$ is the sum
of {the ``kinetic term'' $\sigma E_{A}(k)$ arising from the ABS
  dispersion relation, and the ``potential term'' $-\xi$ resulting
  from dc-voltage biasing.}  {Here, the} $(\xi,k)$ {variables are seen
  as} the {equivalent position-momentum phase-space of a} fictitious
spin-$1/2$ particle. For a given choice of $E$ and $\sigma$,
Eq.~(\ref{key_tilted_band_relation}) defines a curve
$\mathcal{T}_{E,\sigma}$ in {this $(\xi,k)$ plane}, which {we call the
  ``classical trajectory'' in phase space. If} $k$ is used as
parameter, Eq.~(\ref{key_tilted_band_relation}) implies {that
  $\xi(k)$} is simply given by the ABS dispersion relation, up to a
shift of $\xi$ by {$-E$. In} the time-dependent picture{, we have}
$k=\omega_{0}t$, and $\xi$ is a periodic function of time, as expected
for Bloch {oscillations in the solid-state analog
  \cite{Melin2,Melin3}.}

Because of Heisenberg's uncertainty principle, a small bias voltage
induces quantum fluctuations $\Delta \xi \,\Delta k\sim eV$ around the
classical trajectories, and also produces Landau-Zener-Stückelberg
transitions between the two Andreev bands. In the semi-classical
approximation, Landau-Zener tunneling is captured by paths connecting
both classical trajectories $\mathcal{T}_{E,+}$ and
$\mathcal{T}_{E,-}$. Along these tunneling paths, $\xi$ is still a
  real number but $k$ becomes complex, as it is expected for
  evanescent wave-functions in tunneling processes.

\begin{figure}[htb]
  \includegraphics[width=.9\columnwidth]{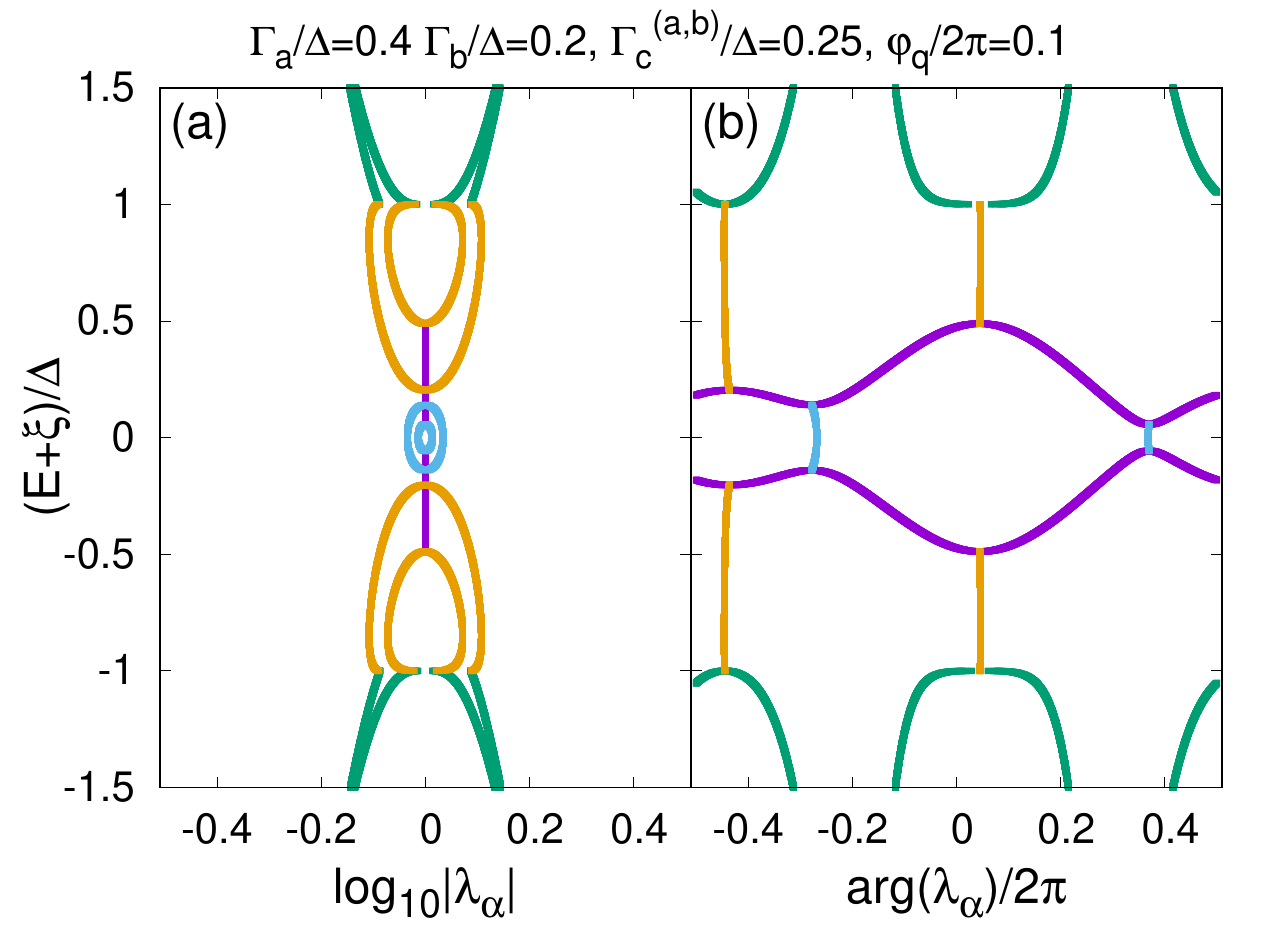}
  \includegraphics[width=.9\columnwidth]{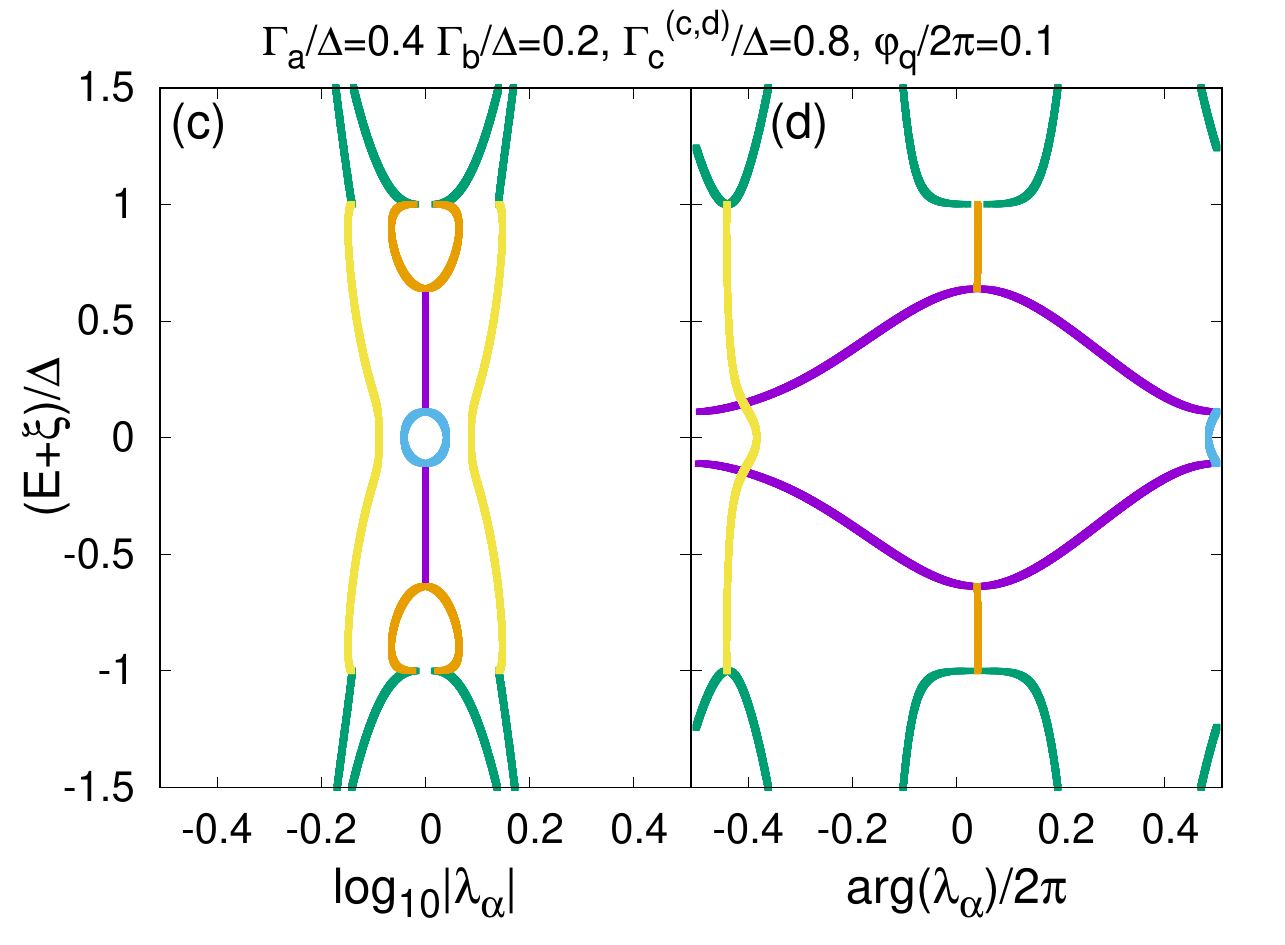}
    \caption{{\it Classical trajectories:} The figure shows the
      classical trajectory $\mathcal{T}_{E,\pm}$, together with the
      tunneling paths, {\it i.e.} the four complex solutions
      $k_\alpha$ of Eq.~(\ref{key_tilted_band_relation}) for a given
      each value of $E+\xi$. {In these plots, we use the complex
      variable $\lambda_{\alpha}=\exp(ik_{\alpha})$.}  
      Panels [(a), (c)] and [(b), (d)] show
      respectively $\log_{10}|\lambda_\alpha|$ and
      $\arg(\lambda_\alpha)/2\pi$ on $x$-axis.
      {The
        $y$-axis on each panel features $E+\xi$ normalized to the gap
        $\Delta$.} The dispersion relation $E_{A}(k)$ (in magenta) has
      two local minima and two local maxima $\mathcal{N}^{(a,b)}=2$ in
      panel (b), as $k$ varies {in the interval
        $-\pi<k<\pi$.} Panel (d) corresponds to a single local minimum
      and maximum $\mathcal{N}^{(c,d)}=1$. The color-code is explained
      in the text.
  \label{fig:semi_class_trajectory}
  }
\end{figure}

The classical trajectories and the tunneling
paths are displayed in Fig.~\ref{fig:semi_class_trajectory}.  Two
representative {sets of parameters are used in}
Figs.~\ref{fig:semi_class_trajectory}~(a)-(b) and
Figs.~\ref{fig:semi_class_trajectory}~(c)-(d),
{differing by the value of} $\Gamma_c/\Delta$ [{\it
    i.e.}  $\Gamma_c^{(a,b)} / \Delta = 0.25$ on panels (a), (b) and
  $\Gamma_c^{(c,d)} / \Delta = 0.8$ on panels (c), (d)], all other
{parameters being the same for all panels} [{\it
    i.e.}  $\Gamma_a/\Delta=0.4$, $\Gamma_b/\Delta=0.2$ and
  $\varphi_q/2\pi=0.1$].  Here, $\Gamma_j=J_j^2/W$ stands from the
contact transparency between the dot and superconducting reservoir
$j$, where $J_j$ is the corresponding tunnel amplitude and $W$ the
band-width. The variable $\varphi_q = \varphi_a + \varphi_b - 2
\varphi_c$ denotes the time-independent quartet phase.
 
On all panels (a)-(d) of Fig.~\ref{fig:semi_class_trajectory}, the
$y$-axis is $E+\xi$ [see Eq.~(\ref{key_tilted_band_relation})]. {On
  $x$-axis, panels (a) and (c) feature $\log_{10}|\lambda_\alpha|$ and
  panels (b) and (d) show $\arg(\lambda_\alpha)/2\pi$,} where
$\lambda_{\alpha}=\exp(ik_{\alpha})$.
The set of solutions to the discrete homogeneous
Eq.~(\ref{homogeneous_difference_eq}) has dimension 4. Then, for each
choice of $\xi$, there are 4 real or complex solutions
$k(\xi)\:\mathrm{mod.}\:2\pi$ of $\mathrm{det}\:L_{0}(\xi,k)=0$.
The magenta curves on
Fig.~\ref{fig:semi_class_trajectory}~(a)-(d) correspond to $\xi$ and
$k$ taking real values, thus with $\log_{10}|\lambda_\alpha|=0$. On
panels (b) and (d), the magenta {data-points coincide with} the ABS
dispersion relations $\pm E_A(k)/\Delta$. {The $|E+\xi|>\Delta$
  branches in green on Figs.~\ref{fig:semi_class_trajectory}~(a)-(d)
  have complex $k$ values, due to the coupling of} the dot level to
the quasi-particle continua above the superconducting gap in the
{leads.} The tunneling paths between the two ABS are shown in
blue. Those between the ABS {and quasiparticle branches} are shown in
orange. The tunneling paths connecting the two continua at energies
$E+\xi<-\Delta$ and $E+\xi>\Delta$ are shown in yellow on panels (c)
and (d).
\begin{figure}[htb]
    \includegraphics[width=\columnwidth]{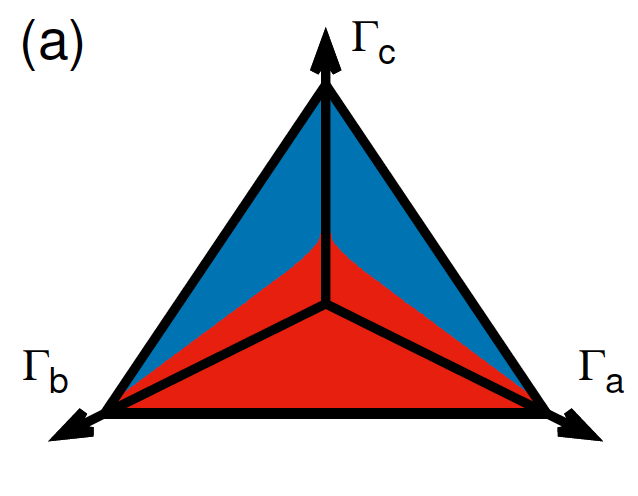}
    \caption{{\it Ternary diagrams for the
        number of minima in $E_{A}(k)$:} the domain in parameter space
      in which the dispersion relation for ${\cal N}=1$ and 2 minima
      is shown in blue and red respectively, for $\varphi_q/2\pi =
      0.2$.}
     \label{fig:ternary_diagram_N}
\end{figure}

The number $\mathcal{N}$ of local minima and maxima in the dispersion
relation $E_{A}(k)$ changes from $\mathcal{N}^{(a,b)}=2$ to
$\mathcal{N}^{(c,d)}=1$ as $\Gamma_c/\Delta$ increases from
$\Gamma_c^{(a,b)}/\Delta$ [panels~(a), (b)] to
$\Gamma_c^{(c,d)}/\Delta$ [panels~(c), (d)].  The values
$\mathcal{N}=1,\,2$ coincide with the number of tunneling {loops}
between the two Andreev bands. Indeed, two or a single tunneling
{loop} can be visualized {in blue} on
Fig.~\ref{fig:semi_class_trajectory}~{(a)} or
Fig.~\ref{fig:semi_class_trajectory}~{(c)}. The jump in $\mathcal{N}$
defines a hypersurface in the $3$-dimensional parameter space
$(\Gamma_{a}/\Gamma_{c},\,\Gamma_{b}/\Gamma_{c},\,\varphi_{q})$,
separating the two regions with ${\cal N}=1,\,2$. A representative
constant-$\varphi_q$ section of this parameter space is shown in
{Fig.~\ref{fig:ternary_diagram_N}.} 
  
  \subsection{$\mathcal{N}=1$ to $\mathcal{N}=2$ transition}
\label{sub_top_phase_transition}

Now we provide more details on the determination of
  $\mathcal{N}$ and of the boundary between $\mathcal{N}=1$ and
  $\mathcal{N}=2$ regions in parameter space.
Eq.~(\ref{explicit_eq_classical_trajectory_2}) shows that it is useful
to specify the variations of $|\Gamma(k)|$ when $k$ varies from $-\pi$
to $\pi$. The Fourier series $|\Gamma(k)|^{2}$ contains harmonics of
the form $e^{imk}$ for $|m|\leq 2$, therefore we have two
possibilities: either $|\Gamma(k)|^{2}$ has two minima and two maxima
in $[-\pi,\pi]$ ($\mathcal{N}=2$), or it has only one minimum and one
maximum ($\mathcal{N}=1$).  In the former case, as $\xi$ increases
across a given classically allowed region, the number of real $k(\xi)$
values is equal to $2$, next $4$, and then $2$ again. An illustration
of this situation is shown on Fig.~\ref{fig:semi_class_trajectory} (b).
In the later case, the number of real $k(\xi)$ values is only 2 throughout each
classically allowed region.  This shows that for any value of $\xi$,
there are at least two $k(\xi)$ values which are not real. Thus, we
obtain a qualitatively simpler situation in which two complex branches
of solutions are decoupled from the real classical trajectory, as
shown by the yellow lines on panels (c) and (d) of Fig.~\ref{fig:semi_class_trajectory}.
 
These two possible regions with $\mathcal{N}=1,\,2$ are separated by a
hypersurface in the 3-dimensional parameter space
$(\Gamma_{a}/\Gamma_{c},\,\Gamma_{b}/\Gamma_{c},\,\varphi_{q})$. Its
equation is obtained by imposing that the first and the second
derivatives of $|\Gamma(k)|^{2}$ vanish simultaneously, which is
equivalent to the vanishing of the resultant for two degree-$4$
polynomials.  A simple geometric interpretation for this hypersurface
can be obtained, in spite of the rather complex corresponding
equation. When $k$ varies, $\Gamma(k)$ moves along an ellipse
$\mathcal{E}$ in the complex plane. An extremum of $|\Gamma(k)|^{2}$
occurs when the origin lies on the normal to $\mathcal{E}$ at the
$\Gamma(k)$ point. At the $\mathcal{N}=1$ to $\mathcal{N}=2$
transition point, a minimum and a maximum of $|\Gamma(k)|^{2}$
collide, so the origin belongs to the intersection of infinitely close
normals, {\it i.e.} it is the curvature center of $\mathcal{E}$ at
$\Gamma(k)$. So the $\mathcal{N}=1$ to $\mathcal{N}=2$ transition
occurs when the origin lies on the evolute of $\mathcal{E}$.

\subsection{Dissipative high energy branches}

Because of the coupling of the dot levels to quasi-particle continua,
we have two complex branches with $\Im k > 0$, which correspond to
quickly decreasing solutions at large $\xi$, and two branches with
$\Im k < 0$, which generate growing solutions. On physical grounds,
the resolvent operator $\mathcal{R}(E)$ will be built from solutions
which decay as $\xi \rightarrow \pm \infty$, so we have to choose the
branches with $\Im k > 0$ as $\xi \rightarrow \infty$ and with $\Im k
< 0$ as $\xi \rightarrow -\infty$. Let us describe the former with the
simplifying assumption that $E+\xi \gg \Delta$. From the explicit form
of $L_{0}$ given in Eq.~(\ref{explicit_L0_above_gap}), we see that, in
this regime, the equation $\mathrm{det}\:L_{0}(\xi,k)=0$ simplifies
and becomes: \be
\exp(2ik)=-\frac{\Delta^{2}\Gamma_{a}\Gamma_{b}}{(E+\xi)^{4}}\exp{\left( i(\varphi_{a}-\varphi_{b})\right)}
,
\ee which leads to \be k=\sigma
\frac{\pi}{2}+\frac{\varphi_{a}-\varphi_{b}}{2}+i\log
\left(\frac{(E+\xi)^{2}}{\sqrt{\Gamma_{a}\Gamma_{b}}\Delta}\right),\;\;\;\;\sigma=\pm
1 .\ee The leading exponential factor in these decaying solutions is:
\be \exp \left\{-\frac{|E+\xi|}{\epsilon}\left(\log
\left(\frac{(E+\xi)^{2}}{\sqrt{\Gamma_{a}\Gamma_{b}}\Delta}\right)-2\right)\right\}
\label{high_energy_decay}
\ee Exactly at the BCS gap, {\it i.e.} if $E+\xi=\pm \Delta$,
Eq.~(\ref{general_eq_classical_trajectory_2}) simplifies into \be
\Gamma(k)\Gamma(k^{*})^{*}=(\sum_{j}\Gamma_{j})^{2}. \ee This equation
has no real $k$ solution unless $\varphi_q=0$, and thus, the vicinity
of the BCS gap then lies in the classically forbidden regions. For further details on the reflections induced at gap edges, see Section I in the Supplemental Material \cite{supplemental}.

\subsection{Non degeneracy condition}
\label{sec:bundle}

Let us consider Eq.~(\ref{order_0_eq_2}). This equation has non
trivial solutions $\chi_{0}(\xi)$ when $(\xi,k=\theta'(\xi))$ lies on
the classical trajectory (extended to complex $k$-values).  A priori,
two cases are possible. Generically, when
$\mathrm{det}\:L_{0}(\xi,k)=0$, the rank of $L_{0}(\xi,k)$ is equal to
unity, so that the direction of the two component spinor
$\chi_{0}(\xi)$ is unambiguously determined. We therefore associate a
line in $\mathbb{C}^{2}$ to each point $(\xi,k)\in \mathbb{R} \times
\mathbb{C}$ such that $L_{0}(\xi,k)$ is of rank 1. A less common
possibility is that the rank of $L_{0}(\xi,k)$ is equal to $0$, {\it
  i.e.} $L_{0}(\xi,k)=0$.  For this to happen, we need to have
simultaneously $E+\xi=0$, and $\Gamma(k)=\Gamma(k^{*})^{*}=0$. Setting
$\lambda=e^{ik}$, this happens when the polynomials
$P(\lambda)=\Gamma_{b}e^{i\varphi_{b}}\lambda^{2}+\Gamma_{c}\lambda+\Gamma_{a}e^{i\varphi_{a}}$
and
$Q(\lambda)=\Gamma_{a}^{*}e^{-i\varphi_{a}}\lambda^{2}+\Gamma_{c}^{*}\lambda+\Gamma_{b}^{*}e^{-i\varphi_{b}}$
have at least one common root. A necessary and sufficient condition
for this to happen is that their resultant $R$ vanishes, which reads
explicitely: \be R =
(\Gamma_{a}^{2}-\Gamma_{b}^{2})^{2}-\Gamma_{c}^{2}(\Gamma_{a}^{2}+\Gamma_{b}^{2}-2\Gamma_{a}\Gamma_{b}\cos\varphi_{q})
= 0. \ee A little algebra shows that $R=0$ is possible in two
situations: either the gap between the two Andreev band closes, or the gap does not
close, but we have $\Gamma_{a}=\Gamma_{b}<\Gamma_{c}/2$ and $\cos\varphi_{q}=1$.
Except for these particular cases, the solutions of Eq.~(\ref{order_0_eq_2})
define a smooth line bundle $\mathcal{B}$ over the classical
trajectory $\mathcal{C}$ (extended to complex k values).

Let us choose a smooth local frame $e(\xi,k(\xi))$ for this bundle,
{\it i.e.} a smooth solution of $L_{0}(\xi,k(\xi))e(\xi,k(\xi))=0$.
To lowest order in the small $\epsilon$ parameter, local semiclassical
solutions have the form $\chi_{0}(\xi)=f(\xi)e(\xi,k(\xi))$ for so far
unknown smooth scalar functions $f(\xi)$, {\it i.e.} they are smooth
local sections of the bundle $\mathcal{B}$. To determine $f(\xi)$
requires more information, which is provided by the first order
equation~(\ref{order_1_eq}).

\subsection{Transport equation}
\label{sec:transport}
To simplify the discussion, we discard the $\xi$-dependence in
$M_{\pm}(\xi)$ by using the approximate forms:
\be
M_{0}(\xi)=\left(\begin{array}{cc} (E+\xi)(1+c) & -\Gamma_{c}\\
  -\Gamma_{c} & (E+\xi)(1+c)
  \end{array}\right)
\label{M_0_simplified}
\ee
\be
M_{+}(\xi)=\left(\begin{array}{cc} 0 &  \Gamma_{b} e^{i\varphi_{b}} \\
  \Gamma_{a} e^{-i\varphi_{a}} & 0
  \end{array}\right)
\label{M_+_simplified}
\ee
\be
M_{-}(\xi)=\left(\begin{array}{cc} 0 & \Gamma_{a} e^{i\varphi_{a}} \\
  \Gamma_{b} e^{-i\varphi_{b}} & 0
\end{array}\right)
\label{M_-_simplified}
\ee
Note that $M_{+}$ and $M_{-}$ are independent of $\xi$ in this approximation.
This is motivated
by the observation that, as explained in subsection~\ref{subsec:ABS_dispersion},
self energies barely affect the shape of the real part of the classical trajectory.  With
this simplification, the $L_{1}(\xi,k(\xi))$ operator is the
following: \be
L_{1}=(-M_{+}e^{ik}+M_{-}e^{-ik})\frac{d}{d\xi}-\frac{i}{2}k'(\xi)(M_{+}e^{ik}+M_{-}e^{-ik})
.
\ee It is convenient to introduce the operator
$K(\xi,k(\xi))=-\frac{i}{2}(M_{+}e^{ik(\xi)}-M_{-}e^{-ik(\xi)})$.
Then, using ${d}/{d\xi}={\partial}/{\partial
  \xi}+k'(\xi){\partial}/{\partial k}$, we have: \be
iL_{1}=2K+\frac{dK}{d\xi}
\label{expression_L_1}
.
\ee
We also have a useful relation between $K$ and $L_{0}$:
\be
\frac{dL_{0}}{d\xi}=(1+c)I+2k'(\xi)K 
\label{K_versus_L_0}
.  \ee Eq.~(\ref{order_1_eq}) imposes that $L_{1}\chi_{0}$ should be
in the image of $L_{0}$, which is of rank 1 on the classical
trajectory. It is convenient to introduce a left eigenvector frame
$\langle e(\xi,k(\xi))|$ such that $\langle e(\xi,k(\xi))|L_{0}(\xi,
k(\xi))=0$. Taking Eq.~(\ref{expression_L_1}) into account, the first
order equation reads \be 2\langle
e|K\frac{d}{d\xi}|\chi_{0}\rangle+\langle
e|\frac{dK}{d\xi}|\chi_{0}\rangle=0
\label{nice_transport_eq}
.
\ee
Geometrically, this defines a connection on the bundle $\mathcal{B}$.
An explicit solution of this equation is derived in Section III of the Supplemental Material \cite{supplemental}. 

\section{Coupled FWS ladders}

\label{subsec_coupled_WS_ladders}

Now, we demonstrate the Bohr-Sommerfeld quantization condition for
periodic orbits. Then, we solve the
Landau-Zener-St\"uckelberg transitions between the Andreev bound state
branches with ${\cal N}=1$ and ${\cal N}=2$ tunneling paths (see Fig.~2 (c)-(d) and Fig.~2
  (a)-(b) respectively).

\subsection{Handling open orbits}
\label{sec:handling}
\begin{figure}[htb]
   \includegraphics[width=\columnwidth]{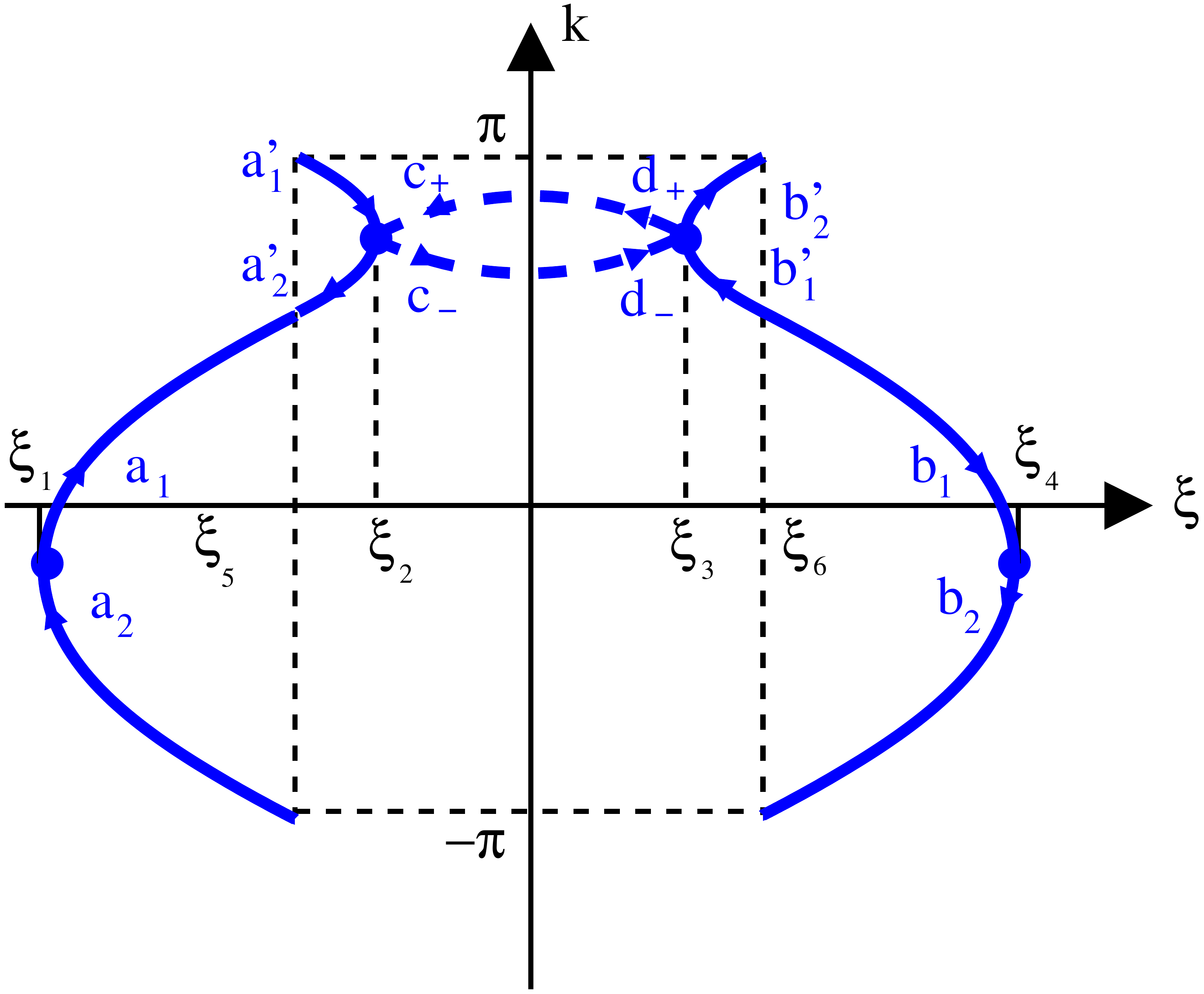}
    \caption{{\it Semiclassical wave-function for $\mathcal{N}=1$}:
      The semiclassical trajectory $\mathcal{T}_{E,-}$
      (resp. $\mathcal{T}_{E,+}$) exhibits turning points at $\xi_{1}$
      and $\xi_{2}$ (resp. $\xi_{3}$ and $\xi_{4}$). The wave vector
      $k(\xi)$ jumps by $2\pi$ at $\xi_{5}$ and $\xi_{6}$. Classical
      trajectories $\mathcal{T}_{E,-}$ and $\mathcal{T}_{E,+}$ are
      depicted by full lines, and they are connected by a pair of
      tunneling paths depicted by dashed lines. Arrows near turning
      points indicate either an increasing phase function
      $\theta(\xi)$ along classical trajectories, or an increasing
      modulus along tunneling paths.
      \label{fig:amplitudes_N=1}
    }
\end{figure}

There is a
domain in parameter space such that the number of real $k(\xi)$ values
is only 2 throughout each classically allowed region, corresponding to
$\mathcal{N}=1$.  The classical orbit is delimited by the two
intervals $\left[\xi_{1},\xi_{2}\right]$ and
$\left[\xi_{3},\xi_{4}\right]$, with $\xi_{2}<\xi_{3}$ (see
Fig.~\ref{fig:amplitudes_N=1}). Note that
$2E+\xi_{1}+\xi_{4}=2E+\xi_{2}+\xi_{3}=0$. When $k$ runs from $-\pi$
to $\pi$, on the left piece of the classical trajectory, $\xi$
decreases from $\xi_{5}$ to $\xi_{1}$, forming the lower part of the
$k_{2}(\xi)$ branch. Then $\xi$ increases from $\xi_{1}$ to $\xi_{2}$,
forming the full $k_{1}(\xi)$ branch. Eventually, $\xi$ decreases from
$\xi_{2}$ to $\xi_{5}$, forming the upper part of the $k_{2}(\xi)$
branch.  Most notations are explained on
Fig.~\ref{fig:amplitudes_N=1}.  The semiclassical Ansatz, to lowest
order in $\epsilon$, amounts to write the wave function in
$\left[\xi_{1},\xi_{2}\right]$ as:
$\Psi(\xi)=\Psi_{1}(\xi)+\Psi_{2}(\xi)$, with: \be
\Psi_{1}(\xi)=a_{1}k'_{1}(\xi)^{1/2}e^{\frac{i}{\epsilon}\int_{\xi_{1}}^{\xi}k_{1}(\xi')d\xi'}e(\xi,k_{1}(\xi))
\label{def_Psi_1}
, \ee where $e(\xi,k(\xi))$ denotes the right zero eigenvector of
$L_{0}(\xi,k(\xi))$ introduced in Section III of the Supplemental Material \cite{supplemental}.
Because we are dealing with an {\em open} classical orbit, the
$k_{2}(\xi)$ branch is discontinuous at $\xi_{5}$, with
$k(\xi+\delta\xi)-k(\xi-\delta\xi)\rightarrow 2\pi$ as
$\delta\xi\rightarrow 0^{+}$, and thus, we prefer to slightly postpone
the discussion of $\Psi_{2}(\xi)$.  As usual, the turning points at
$\xi_{1}$ and $\xi_{2}$ (where two branches meet) need special
care. In the vicinity of $\xi_{1}$, we have $k_{1}(\xi) =
k(\xi_{1})+c(\xi-\xi_{1})^{1/2}+...$, and $\theta(\xi)=
k(\xi_{1})(\xi-\xi_{1})+\frac{2c}{3}(\xi-\xi_{1})^{3/2}+...$, where
$c$ is a positive constant.  As $\xi \stackrel{>}{\rightarrow}
\xi_{1}$, we have: \be \Psi_{1}(\xi) \simeq
\frac{a_{1}}{(\xi-\xi_{1})^{1/4}}e^{\frac{i2c}{3\epsilon}(\xi-\xi_{1})^{3/2}}
\Phi_{1,\mathrm{reg}}(\xi), \ee with $\Phi_{1,\mathrm{reg}}(\xi)$ a
smooth function near $\xi_{1}$.

For $\Psi_{2}(\xi)$, we can use the same definition as
Eq.~(\ref{def_Psi_1}) for $\Psi_{1}(\xi)$, with $k_{1}(\xi)$ replaced
by $k_{2}(\xi)$, as long as $\xi_{1} \leq \xi \leq \xi_{5}$. Near
$\xi_{1}$, we get then: \be \Psi_{2}(\xi) \simeq
\frac{a_{2}}{(\xi-\xi_{1})^{1/4}}e^{-\frac{i2c}{3\epsilon}(\xi-\xi_{1})^{3/2}}
\Phi_{2,\mathrm{reg}}(\xi)
.
\ee The key point here is that
$\Phi_{1,\mathrm{reg}}(\xi_{1})=\Phi_{2,\mathrm{reg}}(\xi_{1})$, so
that, as usual~\cite{Landau}, we can match the various
semiclassical wavefunctions near the turning point at $\xi_{1}$ using Airy
functions.  Imposing decay in the classically
forbidden side $\xi<\xi_{1}$ leads to \be a_{1}=-ia_{2}
\label{matching_tp_1}
.
\ee

Let us for a moment neglect the tunneling processes between the two Andreev bands. We would like to apply a similar relation
for the turning point at $\xi_{2}$. For this, we need the leading behavior of $\Psi_{1}(\xi)$ and $\Psi_{2}(\xi)$ near $\xi_{2}$.
On the one hand, we have
\be
\Psi_{1}(\xi) \simeq \frac{a'_{1}}{(\xi_{2}-\xi)^{1/4}}
e^{\frac{i2c}{3\epsilon}(\xi_{2}-\xi)^{3/2}}\tilde{\Phi}_{1,\mathrm{reg}}(\xi)
,
\ee
where
\be
a'_{1}=a_{1}e^{\frac{i}{\epsilon}\int_{\xi_{1}}^{\xi_{2}}k_{1}(\xi)d\xi}
\label{eq_a'1}
.  \ee For $\Psi_{2}(\xi)$, we have to address the matching problem
across $\xi_{5}$. Let us write \be \Psi_{2}(\xi) =
a_{2}(-k'_{2}(\xi))^{1/2}e^{\frac{i}{\epsilon}\int_{\xi_{1}}^{\xi}k_{2}^{<}(\xi')d\xi'}e(\xi,k_{2}^{<}(\xi))
\ee if $\xi_{1}\leq\xi\leq\xi_{5}$, and \be \Psi_{2}(\xi) =
a'_{2}(-k'_{2}(\xi))^{1/2}e^{-\frac{i}{\epsilon}\int_{\xi}^{\xi_{2}}k_{2}^{>}(\xi')d\xi'}e(\xi,k_{2}^{>}(\xi))
\;\; \ee if $\xi_{5}\leq\xi\leq\xi_{2}$. Here
$k_{2}^{>}(\xi)-k_{2}^{<}(\xi)=2\pi$. This leads to \be
e^{\frac{i}{\epsilon}k_{2}^{>}(\xi_{5})\xi}=e^{\frac{i}{\epsilon}k_{2}^{<}(\xi_{5})\xi}
\ee for $\xi=n\epsilon$, $n$ integer. Then, we obtain \be
a'_{2}=(-1)^{w}a_{2}e^{\frac{i}{\epsilon}\left(\int_{\xi_{1}}^{\xi_{5}}k_{2}^{<}(\xi)d\xi+\int_{\xi_{5}}^{\xi_{2}}k_{2}^{>}(\xi)d\xi+2\pi\xi_{5}\right)}
\label{eq_a'2}
.  \ee Note that $\xi_{5}$ is not necessarily an integer multiple of
$\xi$. It is thus important to keep the last term of the exponential
factor in Eq.~(\ref{eq_a'2}). The winding number $w$ around the origin
of the ellipse described by $\Gamma(k)$ as $k$ increases by $2\pi$. As
explained in the main text, $w=0$ if the origin lies outside the ellipse,
and $w=\pm 1$ if it lies inside. From the expression of $e(\xi,k)$
given in Section III of the Supplemental Material \cite{supplemental}, we get
$e(\xi,k+2\pi)=(-1)^{w}e(\xi,k)$.  With this definition of
$\Psi_{2}(\xi)$, it behaves as follows near $\xi_{2}$: \be
\Psi_{2}(\xi) \simeq \frac{a'_{2}}{(\xi_{2}-\xi)^{1/4}}
e^{-\frac{i2c}{3\epsilon}(\xi_{2}-\xi)^{3/2}}\tilde{\Phi}_{2,\mathrm{reg}}(\xi)
.
\ee As for the turning point near $\xi_{1}$,
$\tilde{\Phi}_{1,\mathrm{reg}}(\xi_{2})=\tilde{\Phi}_{2,\mathrm{reg}}(\xi_{2})$,
so matching with Airy functions leads to \be a'_{1}=-ia'_{2}
\label{matching_tp_2}
.  \ee From the matching conditions Eqs.~(\ref{matching_tp_1}) and
(\ref{matching_tp_2}), and the propagation rules for the amplitudes
      [see Eqs.~(\ref{eq_a'1}), (\ref{eq_a'2})], we obtain the
      Bohr-Sommerfeld quantization condition: \be
      (-1)^{w}e^{\frac{i}{\epsilon}\left(\int_{\xi_{1}}^{\xi_{5}}k_{2}^{<}(\xi)d\xi+\int_{\xi_{5}}^{\xi_{2}}k_{2}^{>}(\xi)d\xi+2\pi\xi_{5}-\int_{\xi_{1}}^{\xi_{2}}k_{1}(\xi)d\xi\right)}=1.
      \ee This can be recast in a much more appealing way, introducing
      $\langle \xi
      \rangle_{\sigma}=\int_{-\pi}^{\pi}\frac{dk}{2\pi}\xi_{\sigma}(k)$,
      where $\xi_{\sigma}(k)$ denotes the piece of the classical
      trajectory such that $\sigma (E+\xi_{\sigma}(k))>0$, $\sigma=\pm
      1$.  Then, the quantization condition becomes: \be
      (-1)^{w}e^{i\frac{2\pi \langle \xi
          \rangle_{\sigma}}{\epsilon}}=1.
\label{1-band_BS_condition}
\ee
The above discussion has considered $\sigma =-1$, but the $\sigma =1$ case is completely analogous.

The solution of Eq.~(\ref{1-band_BS_condition}) reads:
\be
\langle \xi \rangle_{\sigma}=(2n+w)\omega_{0}
\label{quantization_<xi>}
,
\ee
with $n$ arbitrary integer. To go further, it is useful to recall Eq.~(\ref{key_tilted_band_relation}) for
the classical trajectories. It can be recast as
\be
E+\xi_{\sigma}(k)=\sigma E_{A}(k), 
\label{explicit_eq_classical_trajectory_3}
\ee
where $E_{A}(k)$ is positive and $2\pi$ periodic in $k$.
Taking averages over $k$, Eq.~(\ref{quantization_<xi>}) becomes:
\be
E=\sigma \langle E_{A} \rangle - (2n+w)\omega_{0}
\label{simple_WS_ladder}
\ee

This is the
semiclassical form of a single infinite Wannier-Stark ladder, one for
each value of $\sigma$. Using this expression
in~(\ref{explicit_eq_classical_trajectory_3}), we see that
quantization selects an infinite discrete family of classical orbits
given by \be \xi_{\sigma}(k)=\sigma (E_{A}(k)-\langle E_{A}
\rangle)+(2n+w)\omega_{0} .\ee

\begin{figure}[htb]
    \includegraphics[width=.9\columnwidth]{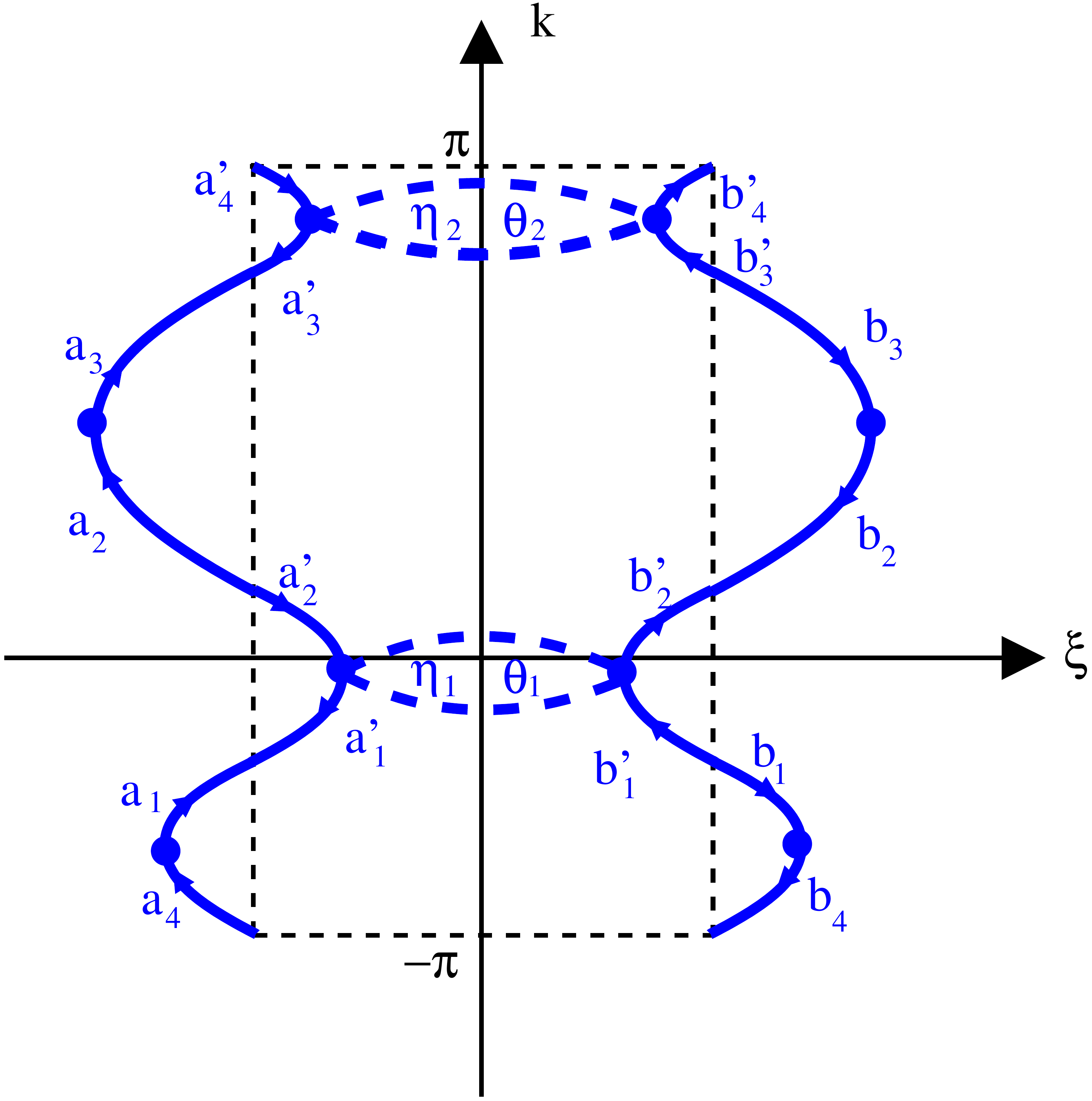}
    \caption{{\it Semiclassical wave-function for $\mathcal{N}=2$}: The semiclassical trajectory $\mathcal{T}_{E,-}$
    (resp. $\mathcal{T}_{E,+}$) exhibits now four turning points. Classical trajectories $\mathcal{T}_{E,-}$ and $\mathcal{T}_{E,+}$ are
    depicted by full lines, and they are connected by a pair of tunneling loops depicted by dashed lines. These tunneling loops are
    characterized by tunneling amplitudes $\eta_{1}$, $\eta_{2}$, and by tunneling phases $\theta_{1}$, $\theta_{2}$.
    Arrows near turning points indicate an increasing phase function $\theta(\xi)$ along classical trajectories.
      \label{fig:amplitudes_N=2}
    }
\end{figure}

\subsection{Tunneling processes: $\mathcal{N}=1$ case}

In the time-dependent picture, Landau-Zener tunneling induces
transitions between the two Andreev levels at any finite voltage. In
the Floquet picture, the effective Hamiltonian becomes
time-independent, but with an additional linear potential $-\xi$. In
the classically forbidden regions $\xi_{2}<\xi<\xi_{3}$, Landau-Zener
transitions are captured by semiclassical solutions associated to
complex $k$-values.  As shown on Fig.~\ref{fig:semi_class_trajectory} (c), there are
two complex $k$ paths connecting the top of the negative energy
Andreev subband to the bottom of the positive energy Andreev
subband. For each $\xi$ such that $\xi_{2}<\xi<\xi_{3}$, the two
values of $k(\xi)$ on these paths are mutually conjugate. Let us
denote by $k_{\tau}(\xi)$ the branch such that the sign of $\Im
(k_{\tau}(\xi))$ is the sign of $\tau=\pm 1$. Then
$k_{+}(\xi)=k_{-}(\xi)^{*}$ when $\xi \in [\xi_{2},\xi_{3}]$.

On this interval, it is then natural to write
$\Psi(\xi)=\Psi_{+}(\xi)+\Psi_{-}(\xi)$, with \be
\Psi_{\tau}(\xi)=c_{\tau}(-i\tau
k'_{\tau}(\xi))^{1/2}e^{\frac{i}{\epsilon}\int_{\xi_{2}}^{\xi}k_{\tau}(\xi')d\xi'}e(\xi,k_{\tau}(\xi))
\label{def_Psi_pm_left}
.
\ee 
Note that $\Psi_{+}(\xi)$ is a semiclassical solution which decreases as $\xi$ moves away from $\xi_{2}$, and which therefore
increases as $\xi$ moves away from $\xi_{3}$.

However, a closer inspection reveals that this form is not correctly
written because one of the components of the local frame
$e(\xi,k_{\tau}(\xi))$ diverges as $\xi+E \rightarrow 0$. Indeed, if
$E+\xi=0$, we have $\rho(k_{+})\rho(k_{-})=0$, with
$\rho(k)=|\Gamma(k)|$. Let us denote by $\tilde{\tau}$ the value of
$\tau$ such that $\rho(k_{\tilde{\tau}})=0$ as $E+\xi=0$.  From the
expressions given in Section III of the Supplemental Material \cite{supplemental} for the local frame
$e(\xi,k_{\tau}(\xi))$, we see that, if the
definition~(\ref{def_Psi_pm_left}) holds when $\xi_{2}<\xi<-E$, then
the smooth solution matching this one at $E+\xi=0$ becomes, when
$-E<\xi<\xi_{3}$: \be \Psi_{\tau}(\xi)=\tilde{\tau}\tau c_{\tau}(i\tau
k'_{\tau}(\xi))^{1/2}e^{\frac{i}{\epsilon}\int_{\xi_{2}}^{\xi}k_{\tau}(\xi')d\xi'}e(\xi,k_{\tau}(\xi))
\label{def_Psi_pm_right}
.
\ee

From Eqs.~(\ref{def_Psi_pm_left}) and (\ref{def_Psi_pm_right}), we
deduce that the connection on the bundle $\mathcal{B}$ has a non
trivial holonomy along the closed path defined by the composition of
the two branches $\xi \rightarrow k_{\tau}(\xi)$, oriented in such a
way that $\xi$ increases from $\xi_{2}$ to $\xi_{3}$ (resp. $\xi$
decreases from $\xi_{3}$ to $\xi_{2}$) when $\tau=1$
(resp. $\tau=-1$). This is a consequence of the presence of an extra
global $\tau$ factor in Eq.~(\ref{def_Psi_pm_right}), which is absent
in Eq.~(\ref{def_Psi_pm_left}). Taking this relative sign into account
, we can write down the Airy matching conditions on both sides of
$\xi_{2}$ as: \be
\frac{c_{-}}{c_{+}}=(2i)\frac{1-(-1)^{w}e^{i\frac{2\pi \langle \xi
      \rangle_{-}}{\epsilon}}}{1+(-1)^{w}e^{i\frac{2\pi \langle \xi
      \rangle_{-}}{\epsilon}}}. \ee Likewise, across $\xi_{3}$, we
get: \be \frac{d_{+}}{d_{-}}=(2i)\frac{1-(-1)^{w}e^{i\frac{2\pi
      \langle \xi \rangle_{+}}{\epsilon}}}{1+(-1)^{w}e^{i\frac{2\pi
      \langle \xi \rangle_{+}}{\epsilon}}},\ee with
$d_{\pm}=\exp(\frac{i}{\epsilon}\int_{\xi_{2}}^{\xi_{3}}k_{\pm}(\xi)d\xi)c_{\pm}$. It
is convenient to write $d_{+}=\lambda c_{+}$ and
$d_{-}=(\lambda^{*})^{-1} c_{-}$ with $|\lambda|<1$.  The
Bohr-Sommerfeld condition takes the form \be
\frac{1-(-1)^{w}e^{i\frac{2\pi \langle \xi
      \rangle_{-}}{\epsilon}}}{1+(-1)^{w}e^{i\frac{2\pi \langle \xi
      \rangle_{-}}{\epsilon}}}\; \frac{1-(-1)^{w}e^{i\frac{2\pi
      \langle \xi \rangle_{+}}{\epsilon}}}{1+(-1)^{w}e^{i\frac{2\pi
      \langle \xi \rangle_{+}}{\epsilon}}} =-\frac{|\lambda|^{2}}{4},
\label{simple_Bohr_Sommerfeld_condition}
\ee where $\lambda$ is the strength of the tunneling amplitude
associated to Landau-Zener-St\"uckelberg processes.
The previous version of the Bohr-Sommerfeld
condition [see Eq.~(\ref{simple_WS_ladder}) above and Eq.~(\ref{adiabatic_Floquet_spectrum}) in the
  article] is recovered in the limit of vanishingly small $|\lambda|$.

Now, tunneling is treated as a small perturbation. We have to
distinguish between the cases of equal or unequal values of $\exp
{\left( {2i\pi \langle \xi \rangle_{+}}/{\epsilon}\right)}$ and $
\exp{\left({2i\pi \langle \xi \rangle_{-}}/{\epsilon}\right)}$. For
equal values, we find \be \delta E_{\sigma}=i\frac{\epsilon
  \lambda^{2}}{4 \pi}\frac{1+(-1)^{w}e^{i\frac{2\pi \langle \xi
      \rangle_{-\sigma}}{\epsilon}}} {1-(-1)^{w}e^{i\frac{2\pi \langle
      \xi \rangle_{-\sigma}}{\epsilon}}}
\label{delta_E_non_degenerate}       
, \ee where $\delta E_{\sigma}$ is real-valued. The wave function is
mostly localized on the $\sigma$ piece of the classical trajectory.

The situation changes qualitatively in the degenerate case. The degeneracy is lifted at first order in $|\lambda|$ according to
\be
\delta E_{\sigma}=\pm \sigma \frac{\epsilon}{2\pi}|\lambda|
\label{delta_E_degenerate}
.  \ee The right-hand side of Eq.~(\ref{delta_E_degenerate}) is much
larger than its counterpart in Eq.~(\ref{delta_E_non_degenerate}) at
small $|\lambda|$. This is the analogue of energy level repulsion, in
the setting of Floquet theory for time periodic Hamiltonians.  This
phenomenon has already been reported in our previous numerical
study~\cite{Melin2}.

The support of the semiclassical wave function is very different,
depending on whether the uncoupled Wannier-Stark ladders are distinct
or degenerate. In the nondegenerate case, the solutions are strongly
localized on one piece $\sigma$ of the classical trajectory. In the
degenerate case, they are linear superpositions with equal weights of
semiclassical wave functions associated to {\em both} pieces of the
classical trajectory. These superpositions appear clearly in the
resolvent when the two Wannier-Stark ladders are nearly degenerate, as
shown in Fig.~2 of the Supplemental Material \cite{supplemental}.

\subsection{Tunneling processes: $\mathcal{N}=2$ case}

Let us now consider the case when the function $\xi(k)$ on either piece of
the classical trajectory has two minima and two maxima when $k$
increases from $-\pi$ to $\pi$, {\it i. e.} $\mathcal{N}=2$.  As shown
on Fig.~\ref{fig:semi_class_trajectory} (a), the two classically allowed regions are
now connected by two tunneling loops. More of the notations used here are
shown in Fig.~\ref{fig:amplitudes_N=2}.  Imposing the Airy
matching rules at each of the two turning points located at the
extremities of the tunneling loop gives: \be
\left(\begin{array}{c}a'_{2}\\ b'_{2}\end{array}\right)=S_{1}\left(\begin{array}{c}a'_{1}\\ b'_{1}\end{array}\right),\;\;\;\;
\left(\begin{array}{c}a'_{4}\\ b'_{4}\end{array}\right)=S_{2}\left(\begin{array}{c}a'_{3}\\ b'_{3}\end{array}\right)
,\ee with: \be S_{j}=i \left(\begin{array}{cc} (1-\eta_{j}^{2})^{1/2}
  & \eta_{j}e^{-i\theta_{j}} \\ \eta_{j}e^{i\theta_{j}} & -
  (1-\eta_{j}^{2})^{1/2}\end{array}\right) \ee Denoting by
$k_{j,+}(\xi)$ the tunneling branch with a positive imaginary part for
$k$, we set $\lambda_{j}=\exp(\frac{i}{\epsilon}\int
k_{j,+}(\xi)\:d\xi)$, where the integral is taken on the $j$-th
tunneling path ($j=1,2$), and $|\lambda_{j}|\ll 1$ in the small
voltage limit. Then, the parameters entering the unitary matrix
$S_{j}$ are: \be
\eta_{j}=\frac{|\lambda_{j}|}{1+\frac{|\lambda_{j}|^{2}}{4}},\;\;\;\;e^{i\theta_{j}}=\tilde{\tau}_{j}\frac{\lambda_{j}}{|\lambda_{j}|}
,\ee where $\tilde{\tau}_{j}=\pm 1$.  Between the two tunneling loops, we
have the usual semiclassical propagation of amplitudes: \be
\left(\begin{array}{c}a'_{3}\\ b'_{3}\end{array}\right)=P_{1}\left(\begin{array}{c}a'_{2}\\ b'_{2}\end{array}\right),\;\;\;\;
\left(\begin{array}{c}a'_{4}\\ b'_{4}\end{array}\right)=P_{2}\left(\begin{array}{c}a'_{3}\\ b'_{3}\end{array}\right)
,\ee with: \be P_{j}=-i \left(\begin{array}{cc} e^{i\varphi_{jL}} & 0
  \\ 0 & - e^{i\varphi_{jR}} \end{array}\right). \ee The phase factors
$e^{i\varphi_{jL}}$ and $e^{i\varphi_{jR}}$ are expressed in terms of
oscillating integrals of the form $\exp(\frac{i}{\epsilon}\int
k(\xi)\:d\xi)$ taken on appropriate paths. When $j=2$, an extra Berry
phase factor $(-1)^{w}$ has to be taken into account. Setting
$M=P_{2}S_{2}P_{1}S_{1}$, the Bohr-Sommerfeld quantization condition
reads: \be \mathrm{det}(M-I)=0
\label{242_Bohr_Sommerfeld_condition}
.
\ee
The entries of $M$ are:
\begin{eqnarray*}
  M_{11} & = & (1-\eta_{1}^{2})^{1/2}(1-\eta_{2}^{2})^{1/2}e^{i\varphi_{L}} - \eta_{1}\eta_{2}e^{i(\varphi_{1R}+\varphi_{2L}+\theta_{1}-\theta_{2})} \\
  M_{22} & = & (1-\eta_{1}^{2})^{1/2}(1-\eta_{2}^{2})^{1/2}e^{i\varphi_{R}} - \eta_{1}\eta_{2}e^{i(\varphi_{1L}+\varphi_{2R}-\theta_{1}+\theta_{2})} \\
  M_{12} & = & \eta_{1} (1-\eta_{2}^{2})^{1/2}e^{i(\varphi_{L}-\theta_{1})} + \eta_{2} (1-\eta_{1}^{2})^{1/2}e^{i(\varphi_{1R}+\varphi_{2L}-\theta_{2})} \\
- M_{21} & = & \eta_{1} (1-\eta_{2}^{2})^{1/2}e^{i(\varphi_{L}+\theta_{1})} + \eta_{2} (1-\eta_{1}^{2})^{1/2}e^{i(\varphi_{1L}+\varphi_{2R}+\theta_{2})}
.
\end{eqnarray*}
Here, we have introduced $\varphi_{L}=\varphi_{1L}+\varphi_{2L}$,
$\varphi_{R}=\varphi_{1R}+\varphi_{2R}$. If we shift the energy $E$ by
$\delta E$, Eq.~(\ref{key_tilted_band_relation}) shows that the
classical trajectory is shifted along the $\xi$ axis by
$\delta\xi=-\delta E$.  A simple analysis (using integration by parts)
of the phase factors involved in the entries of $M$ shows that $M$ is
multiplied by $\exp(i\frac{2\pi \delta E}{\epsilon})$. This implies
that the Bohr-Sommerfeld quantization condition is invariant when $E$
is shifted by integer multiples of $\epsilon=2eV/\hbar$, and therfore,
we get a periodic Wannier-Stark ladder spectrum.

Let us now treat tunnel amplitudes $\eta_{1}$,  $\eta_{2}$, as small perturbations. When these amplitudes vanish ({\it e.g.} as the bias voltage
$V \rightarrow 0$), Eq.~(\ref{242_Bohr_Sommerfeld_condition}) becomes $(e^{i\varphi_{L}}-1)(e^{i\varphi_{R}}-1)=0$, and we have two uncoupled ladders,
one associated to each piece of the classical trajectory. When we switch on small tunneling amplitudes, we have to distinguish between the non-degenerate case $e^{i\varphi_{L}} \neq e^{i\varphi_{R}}$
  and the degenerate one. In the former case, the energy shift $\delta E$ for the left ladder (we assume $e^{i\varphi_{L}}=1$) is given by:
\begin{eqnarray}
  \frac{2\pi \delta E}{\epsilon}&=&\frac{\eta_{1}^{2}+\eta_{2}^{2}}{2}\cot(\frac{\varphi_{R}}{2})\\
  \nonumber
&&+\eta_{1}\eta_{2}
\frac{\cos(\varphi_{1L}+\frac{\varphi_{2R}-\varphi_{1R}}{2}-\theta_{1}+\theta_{2})}{\sin(\frac{\varphi_{R}}{2})}
\label{delta_E_non_degenerate_242}
\end{eqnarray}
The energy shift for the right ladder is given by a similar expression, after replacing $R$ by $L$.
In the degenerate case ($e^{i\varphi_{L}}=e^{i\varphi_{R}}=1$), the degenerate levels are repelled from each other
according to:
\begin{eqnarray}
\label{delta_E_degenerate_242}
  \frac{2\pi \delta E}{\epsilon}&=&\pm\left(\eta_{1}^{2}+\eta_{2}^{2}+\right.\\
  &&\left.+2\eta_{1}\eta_{2}
  \cos(\varphi_{1L}+\varphi_{2R}-\theta_{1}+\theta_{2})\right)^{1/2}.
  \nonumber
\end{eqnarray}
Here, the new qualitative feature is the presence of interferences between
the two tunneling paths.  They appear {\it via} the voltage-dependent phases
$\theta_{1}$ and $\theta_{2}$ in Eqs.~(\ref{delta_E_non_degenerate_242}),
(\ref{delta_E_degenerate_242}).  For an illustration of such
interferences, see Fig.~4 (d) in Ref.~\onlinecite{Melin3}.


\end{document}